\nonstopmode
\pdfoutput=1
\documentclass[preprint]{aastex}
\usepackage{epsfig}

\slugcomment{Accepted by ApJ, 2016 February 5}
\shorttitle{AGN Variability: Propagating Turbulent Jets}
\shortauthors{Pollack, Pauls and Wiita}

\begin{document}
\title{Variability in Active Galactic Nuclei from Propagating \\ Turbulent Relativistic Jets}

\author{Maxwell Pollack\altaffilmark{1}, David Pauls and Paul J.\ Wiita}

\affil{Department of Physics, The College of New Jersey}
\affil{P.O.~Box 7718, Ewing, NJ 08628-0718, USA}
\email{wiitap@tcnj.edu}

\altaffiltext{1}{current address: Department of Astronomy, University of Wisconsin-Madison, 475 N.\ Charter St., Madison, WI  53706-1507, USA}

\begin{abstract}
	We use the Athena hydrodynamics code to model propagating two-dimensional relativistic jets as approximations to the growth of radio-loud active galactic nuclei for various input jet velocities and jet-to-ambient matter density ratios. Using results from these simulations we estimate the changing synchrotron emission by summing the fluxes from a vertical strip of zones behind the reconfinement shock, which is nearly stationary, and from which a substantial portion of the  flux variability should arise.  We explore a wide range of timescales by considering two light curves from each simulation; one uses a relativistic turbulence code with bulk velocities taken from our simulations as input, while the other uses the bulk velocity data to compute fluctuations caused by variations in the Doppler boosting due to changes in the direction and the speed of the flow through all zones in the strip.  We then calculate power spectral densities (PSDs) from the light curves for both  turbulent and bulk velocity origins for variability.  The range of the power-law slopes of the PSDs for the turbulence induced variations is $-1.8$ to $-2.3$, while for the bulk velocity produced variations this range is $-2.1$ to $-2.9$; these are in agreement with most observations.  When superimposed, these power spectra span a very large range in frequency (about five decades), with the turbulent fluctuations yielding most of the shorter timescale variations and the bulk flow changes dominating the longer periods.
\end{abstract}

\subjectheadings{galaxies: active --- BL Lacertae objects: general --- galaxies: jets --- quasars: general }

\section{Introduction}

Variability in observed emission can be considered a defining characteristic of active galactic nuclei (AGNs), and for the roughly 10\% of AGNs that are radio-loud (e.g., Jiang et al.\ 2007) the majority of this variable emission is understood to arise from the relativistic flows of plasma along two oppositely directed jets (e.g.\ Urry \& Padovani 1995).  When viewed at small angles to the jet direction the Doppler boosting makes the emission from the approaching jet appear dramatically brighter and also shortens the observed timescales with respect to those in the emitted frame, thereby explaining many of the properties of blazars (e.g.\ Blandford \& Rees 1978; Lister 2001; Gopal-Krishna et al.\ 2003).  Multiband radio studies and very long baseline interferometry (VLBI), coupled with theoretical models, have provided extremely strong evidence for the presence of both moving and standing shocks in these jets (e.g., Marscher \& Gear 1985; Hughes et al.\ 1991; Lister et al.\ 2001, 2009; Marscher et al.\ 2008, 2010),  indicating significant changes in the fluid flow and/or the density of material ejected into the jets;  it is now  accepted that the largest flares arise from the production and relativistic propagation of new components seen as radio knots.  
Changes in the overall direction of the inner portions of the jet, or at least its brightest portions, have also been demonstrated via VLBI (e.g.\ Biretta et al.\ 1986; Ros et al.\ 2000; Piner et al.\ 2003; Caproni \& Abraham 2004). Even modest changes in direction  (e.g.\ Camenzind \& Krockenberg 1992; Gopal-Krishna \& Wiita 1992) have long been recognized as one way to produce significant changes in the flux and polarization (e.g.\ Gopal-Krishna \& Wiita 1993; Piner et al.\ 2008).   It is to be expected that turbulence is produced within these jets, at least in the vicinity of shocks, and thus some of the  variations should arise from such smaller scale motions as has been suggested theoretically (Marscher \& Travis 1991;  Marscher 2014; Calafut \& Wiita 2015) and strongly
supported by observations of blazars (e.g.\ Marscher et al.\ 2008, 2010; Bhatta et al.\ 2013).  In addition, there is the possibility that portions of the jet are moving much faster than other portions, and such  misaligned mini-jets  could also produce some extremely rapid fluctuations (Giannios et al.\ 2009; Biteau \& Giebels 2012).    Variability on a wide range of timescales can  be produced within the  accretion disks feeding the central black holes (e.g.\ Czerny 2006).  This presumably dominates the variations from radio-quiet AGNs though not those of radio-loud ones because the special relativistic boosting of the jet emission is so important in the latter (Urry \& Padovani 1995). Some of the variations in the jet emission might be traced to plasma fluctuations in the disk being advected into the base of the jets (e.g.\ Wiita 2006) but the exact origins of the initial fluctuations are not addressed in this work.  Here we address the question of whether variations in the bulk flow of a propagating relativistic hydrodynamic (RHD) jet along with sub-grid mildly relativistic turbulence can produce light curves and power spectra resembling those of radio-loud AGNs.

				Hydrodynamical simulations of propagating jets are of critical importance to the understanding of radio galaxies and have a long history (e.g., Wiita 1978; Norman et al.\ 1982).   Although most of the understanding of two- and three-dimensional (2D and 3D) flows was achieved by the mid-1990s (e.g.\ Hardee \& Norman 1988, 1990; Burns et al.\ 1991; Hooda et al.\ 1994; Bassett \& Woodward 1995; Bodo et al.\ 1995), only with superior computing powers could hydromagnetic flows be properly considered (see the review by Ferrari 1998).  Continuing these studies remains important because they allow us to model phenomena that occur naturally in the flow of the jets and understand how they affect the light curves and power spectra in different bands. In most instances, there is insufficient angular resolution in even the best VLBI imaging to show many details within the jets, so an understanding of the traces that factors such as turbulence and variations in jet velocity and density leave on light curves is needed.  In this paper we focus on producing rough models of key phenomena that occur naturally in the flows of relativistic jets; the unique feature of this work is our combining variations due to irregularities in the bulk properties of propagating jet flows with those arising from turbulence.
			
				The early simulations of non-relativistic jets alluded to above were followed by simulations of special relativistic jets.  These were carried out by several groups, with some focusing on the propagation on small scales corresponding to the parsec-scale structures revealed by VLBI (e.g.\ Mart{\'i} et al.\ 1995, 1997; van Putten 1996;  Mioduszewski et al.\ 1997), while others focused on propagation out to larger scales and production of the structures reflected on the multi-kiloparsec scale of powerful radio galaxies (e.g.\ Rosen et al.\ 1999; Hardee et al.\ 2001; Carvalho \& O'Dea 2002a,b; Leismann et al.\ 2005; see review by Mart{\'i} \& M{\"u}ller 2003).  Advances in computing power  made since then allow us to run simulations with comparable or higher resolution and accuracy on  personal machines that were only possible on supercomputers of the time.  Also, there have been developments in computational fluid dynamics (CFD) software that minimize numerical error, maximize efficiency, and allow the inclusion of more advanced physics.
An excellent astrophysical CFD code is Athena, developed by J.\ Stone and colleagues (Gardiner \& Stone 2005; Stone et al.\ 2008; Beckwith \& Stone 2011).  Athena is a highly efficient, grid-based magnetohydrodynamical (MHD) code that possesses advanced capabilities such as self-gravity, viscosity, and (particularly important for our purposes) special relativity.  It is well documented and relatively easy to implement and has allowed us to conduct true RHD simulations of radio jets (albeit not MHD ones to date)  at both reasonably high resolution and for the high Lorentz factors that better approximate the jets of actual blazars.
								
				 The number of grid zones spanning the jet diameter in our simulations (and the vast majority of earlier simulations discussed above) is typically around 20, while the turbulence within a jet will normally be produced on scales smaller than such zones.   Hence, a separate sub-grid turbulence model must be introduced; in a first approximation such models will be characterized by a fixed maximum turbulent velocity and the total energy injected into the turbulence.  We modeled turbulence in the jet flow following Calafut \& Wiita (2015); their approach approximates relativistic turbulence through refinement of larger eddies into smaller eddies contained within the larger ones. We take the largest zones in which turbulence occurs to correspond to the smallest grid zones in our bulk velocity simulations, which, in our scenario, cannot resolve the turbulence.   Each level of smaller eddies possess lower speeds; for a non-relativistic Kolmogorov turbulent spectrum, the speeds scale as the one-third power of the eddy size in the inertial range.  In this model, each eddy at each level of refinement is assigned an initial random direction with respect to the jet axis and these changing velocities determine variable Doppler boosting factors for each portion of the flow and thus preferentially instantaneously beam their emissions strongly in those directions.  A simplification in this work is the assumption that all of the turbulent motion is in a fixed plane parallel to the jet axis and hence it is only a 2D approximation to the full 3D turbulence.  Turning this into a 3D turbulent model would require adding another random variable for the orientation of the eddy with respect to the turbulent cell; however, this would make negligible difference to the output light curves, while significantly increasing the computational complexity.  While not a true simulation of turbulent flow, this method did model RHD turbulence with reasonable accuracy, as evidenced  both by the shapes of the light curves and the power spectral density (PSD) slopes (Calafut \& Wiita 2015, hereafter CW).  This paper also compared results from the standard Komolgorov turbulent spectrum with a well motivated special relativistic turbulence spectrum from Zrake \& MacFadyen (2013) that ties the Lorentz factor to the length scale $\ell$. We adopted this relativistic turbulence spectrum, as discussed further in  \S 3.2. 
				
				Our work bears a resemblance to the recent work of Marscher (2014) in that it calculates fluxes from a standing shock for a portion of a relativistic turbulent hydrodynamic jet. Marscher's turbulent, `extreme multi-zone' model provides a more accurate and sophisticated method for calculating these fluxes and has the major advantage of computing fluxes and polarizations at a wide range of frequencies. Nonetheless, our model has some advantages that complement Marscher's work. An important difference in the models is that we calculated separate light curves for both shorter timescale turbulent variability and longer timescale bulk variability in the hope of creating a physically reasonable yet more expansive look at the power spectrum, which spans about five decades for our simulations. Marscher (2014) assumes a constant bulk velocity as the input for the calculation of the variability arising from the turbulent cells whose maximum velocities are also taken as input variables, and states that the value for the bulk velocity should be taken from relativistic MHD simulations. Although our simulations are not MHD, they are RHD and propagating. Our bulk velocities change with time and differ across the width of the flow, so the shear between zones provides a natural source for the maximum speeds of the sub-grid turbulence and  thus  provides a useful step toward using differences between our more accurate bulk velocity input data to feed into subsequent, more precise, sub-grid turbulent variability calculations. Neither Marscher's nor our methods of incorporating turbulence are true simulations of turbulent flow throughout the inertial range, but the CW approach implements a relativistic turbulence calculation, instead of using the standard Kolmogorov spectrum. Most likely, however, this will barely affect the simulations, as CW found that highly relativistic turbulence does not resemble data obtained from observing active relativistic jets, and therefore the turbulence in these jets is most likely minimally or moderately relativistic and almost certainly subsonic (i.e., the maximum turbulent velocity should be below $3^{-1/2} c$).

In \S 2 we provide a description of the RHD simulations while \S 3 provides an analysis of the variability produced by both the large scale bulk velocity changes and the smaller scale turbulent motions.  Our results are summarized in \S 4 and our conclusions are given in \S 5.
				
				\section{RHD Simulations}

				Our simulations of RHD jets were accomplished using the Athena code  (Gardiner \& Stone 2005; Stone et al.\ 2008; Beckwith \& Stone 2011).  Athena is a grid-based MHD code with the capability to include special relativity and static mesh refinement, which make it a very attractive choice for use in simulating astrophysical jets. Athena works by numerically solving a system of partial differential equations for conservation of mass, momentum, magnetic flux, and  energy, along with the equation of state at  successive timesteps for each zone on a grid of customizable size. These equations are solved for density, pressure, velocity, internal energy, and magnetic field. Since we performed RHD simulations, not MHD ones, the magnetic field was set to zero. 

							
				
				
				

				 A predesigned hydrodynamic jet problem is available in the problem files bundled with Athena\footnote{https://trac.princeton.edu/Athena}.  It requires an input jet velocity and jet-to-ambient matter density ratio, and  we modified this initial setup for our purposes. Athena permits the user to optimize the simulation by allowing a choice of the order of reconstruction, as well as providing options such as first-order flux correction, static mesh refinement, a selection of boundary condition (BC) types, and the ability to specify the Courant-Friedrichs-Lewy (CFL) number. 
				
First-order flux correction alters the values of parameters if these values overstep their allowed bounds; for matter density and velocity these are zero and the speed of light, respectively. However, first-order flux correction fails and the run terminates if repeated impossible values continue to be returned by the program.  We determined that first-order flux correction provided a crucial tool that extended the lifetime of many of our runs.  Static mesh refinement alters the resolution of a rectangular section of the grid by a factor of two per refinement level. This allows more focus on parts of the grid that contain interesting phenomena, such as the jet or shocks. However, despite the obvious advantage of this approach we found that the code often terminated prematurely when static mesh refinement was applied in the RHD version of the code.  As we had reasonably adequate resolution in our production runs across the slab jet width ($\sim 20$ zones) even without static mesh refinement, we chose not to employ it.  
						
				Aside from the ``nozzle'' on the left side of the grid where the jet is introduced, where an inflow BC is obviously required, the BCs on the top, bottom, and right sides should be set to outflow so that the matter pushed to the boundary does not reflect and inappropriately interfere with the jet. However, doing so means that we are losing matter from the grid at late stages, which affects the jet evolution and reduces the accuracy of our results once the flow reaches the length of the grid. We believe, however, that the jet structure is only mildly impacted by this loss.  This is particularly true for our main interest here, the inner region of the jet near the reconfinement shock, which is far removed from those grid edges.  Hence we do sometimes continue the analysis of runs even after the bow shock and the end of the jet have completely crossed the grid.  We considered both outflow and reflection BCs at the left boundary (aside from the inflow region).  The reflection BC choice has some attractiveness since a real system would have a pair of oppositely directed flows whose backflows would collide in that region. However, after conducting runs with both reflection and outflow BCs we decided that the reflected matter interfered too much with the inner part of the cocoon and the jet itself, resulting in instabilities being amplified; hence our production runs used outflow BCs above and below the jet inlet region.

				The CFL number, the factor multiplying the ratio of the grid size divided by the maximum velocity that determines the computational timestep, is a limiting factor for both the accuracy and computation speed of the simulation. If the CFL number is not set low enough, the solutions to the fluid motion differential equations will not converge correctly, and the simulation results will not be accurate. However, the lower the CFL number, the longer the computation time. With most of the lower-velocity jets, stability was lost very early on due to slight perturbations arising from numerical instabilities if the CFL number was 0.2 or greater.  Hence we chose a CFL number of 0.1 for our production runs, which increased the computation time but in turn allowed most of the jets to remain stable until they had traversed at least a large majority of the grid. 
				
				\begin{figure}
						\begin{center}
								\begin{tabular}{c c}
										\epsfig{file=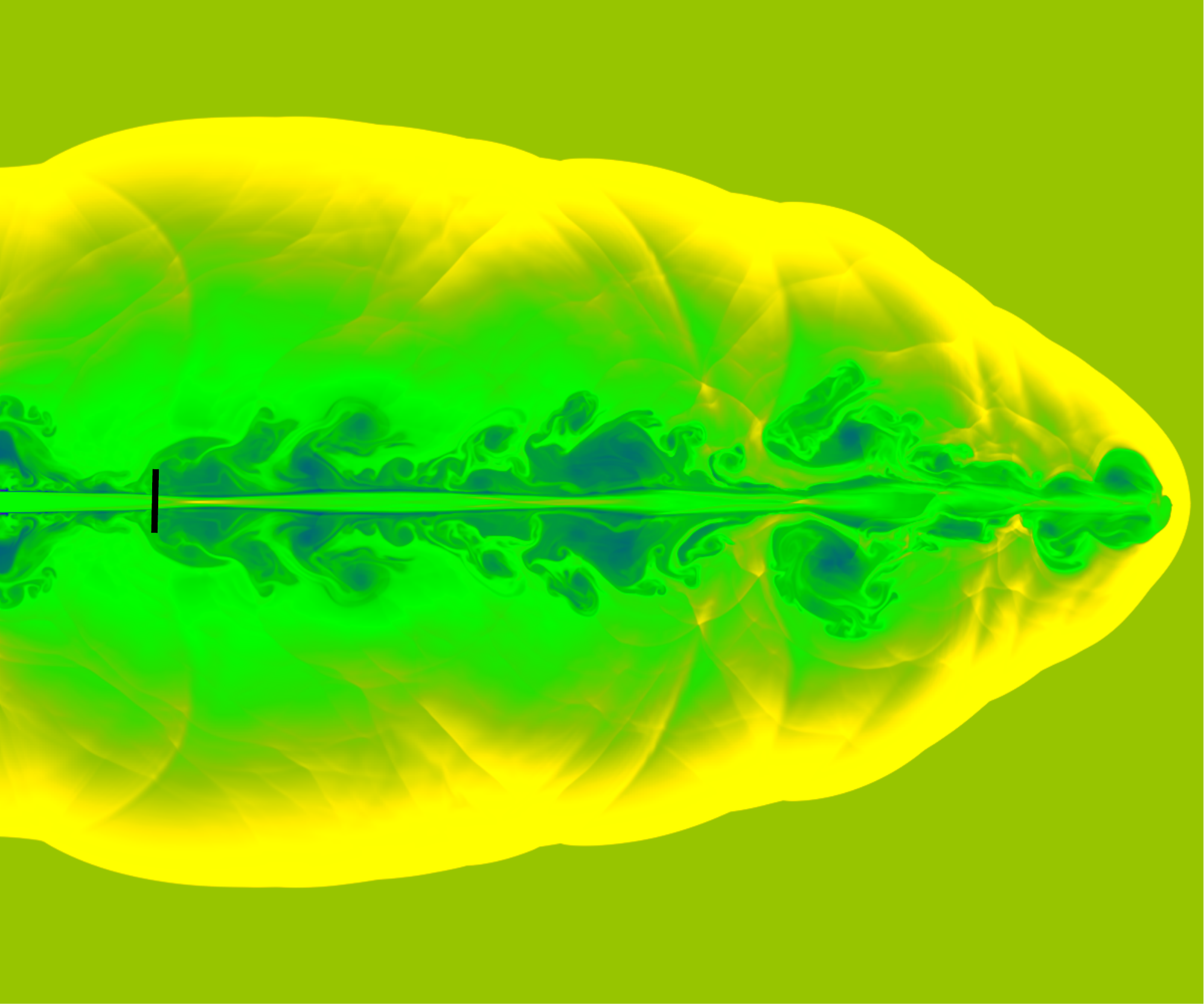, height=2.4 in} & \epsfig{file=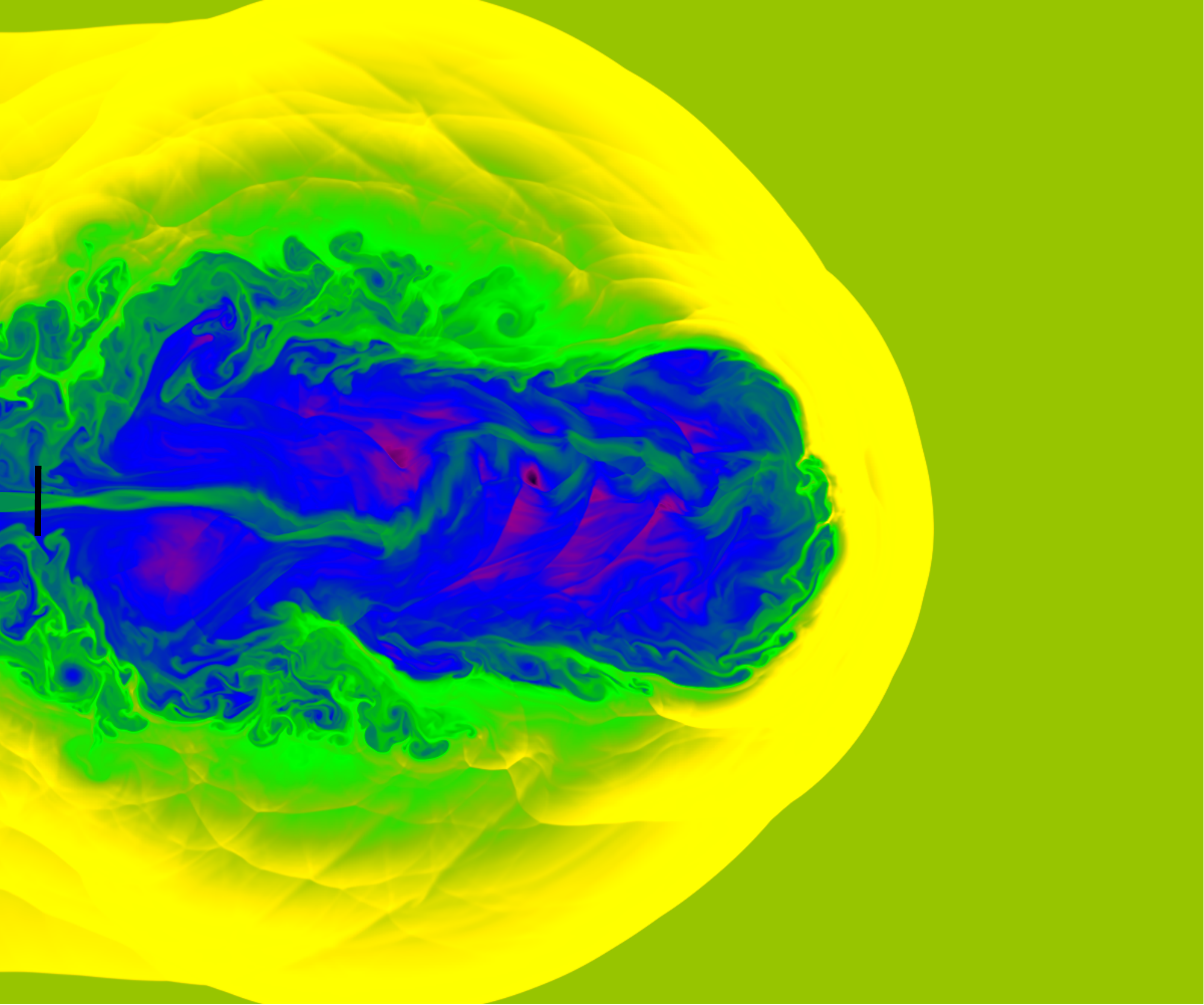, height= 2.4 in} \\
										\epsfig{file=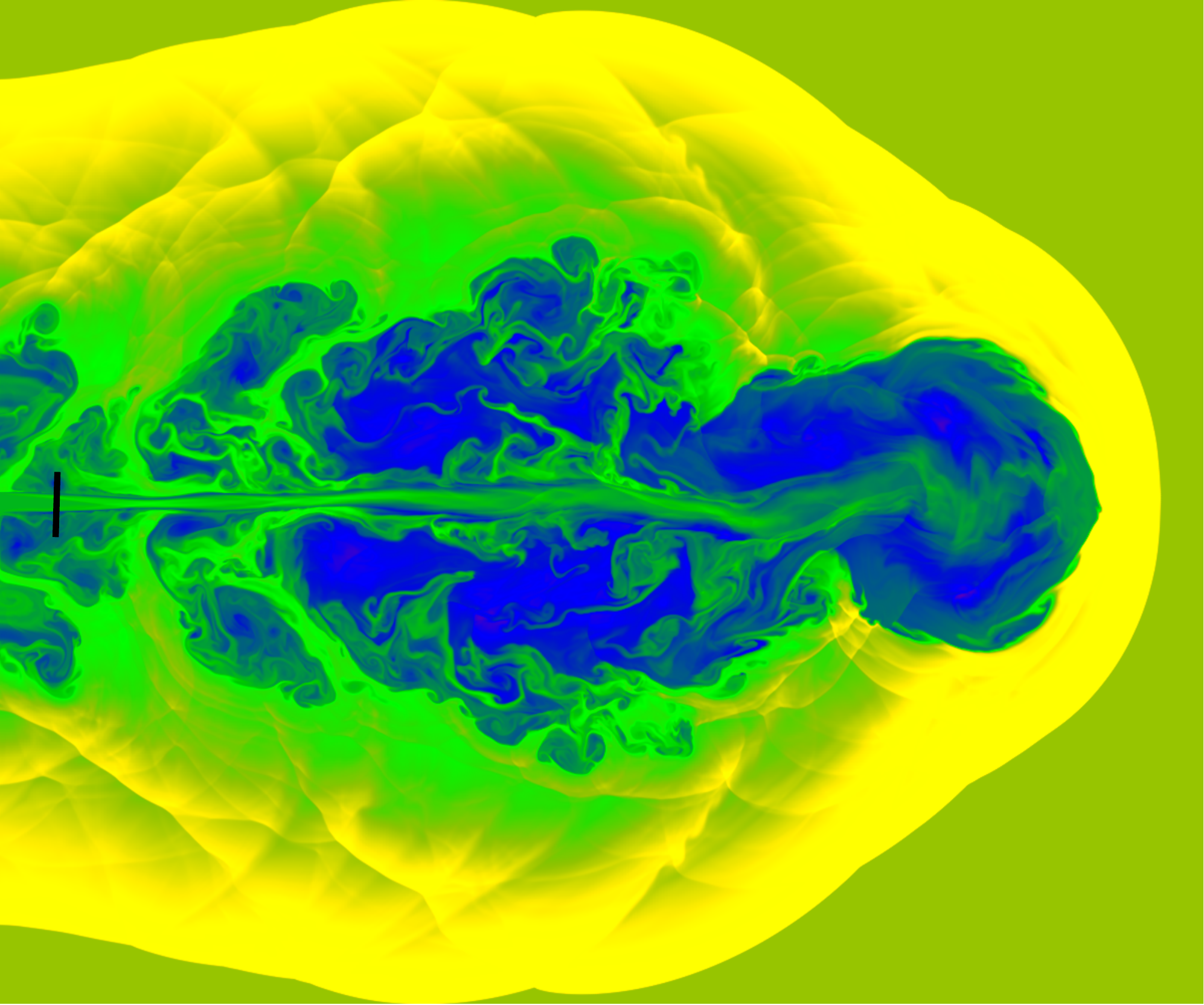, height=2.4 in} & \epsfig{file=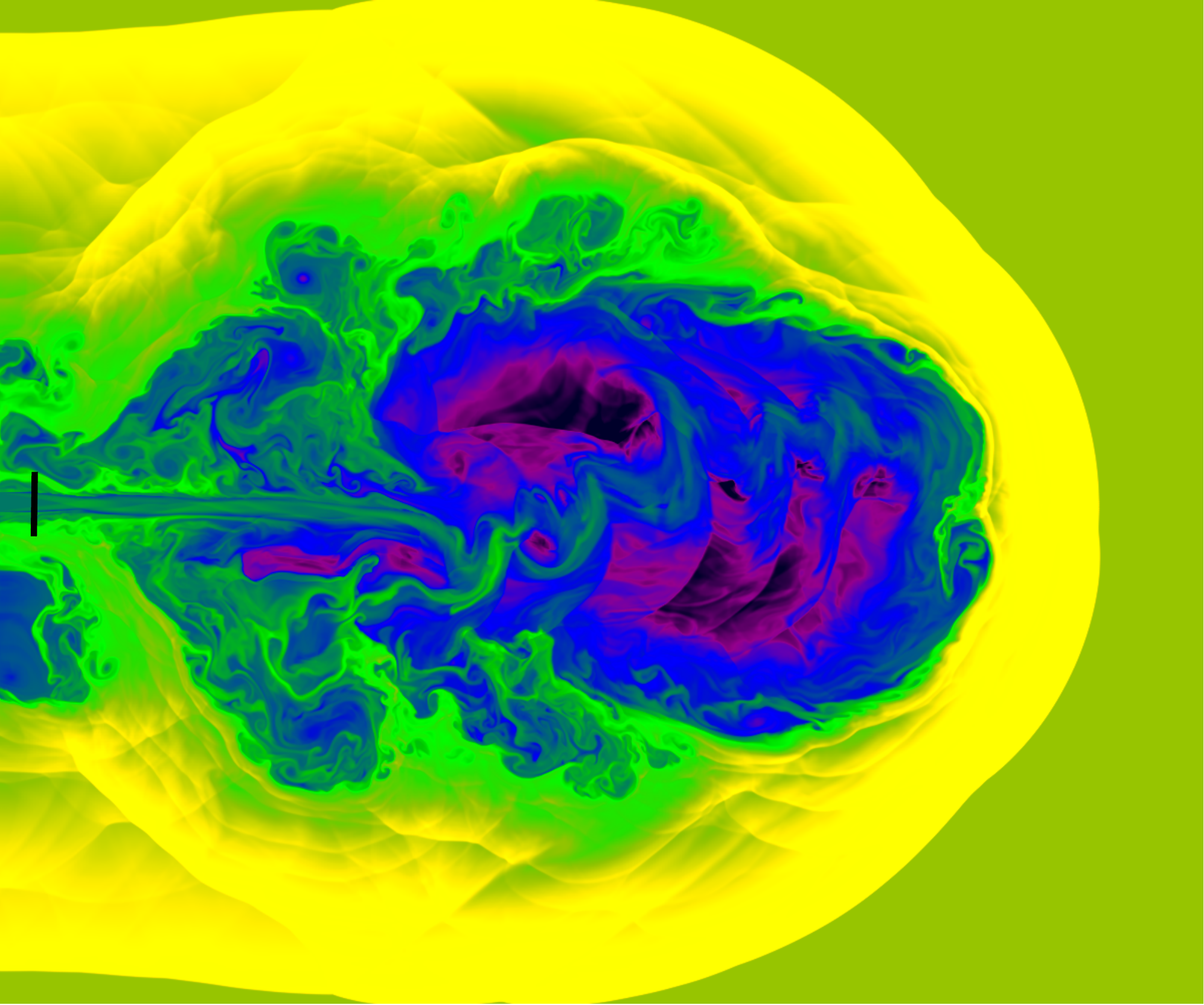, height=  2.4 in} \\	
										\epsfig{file=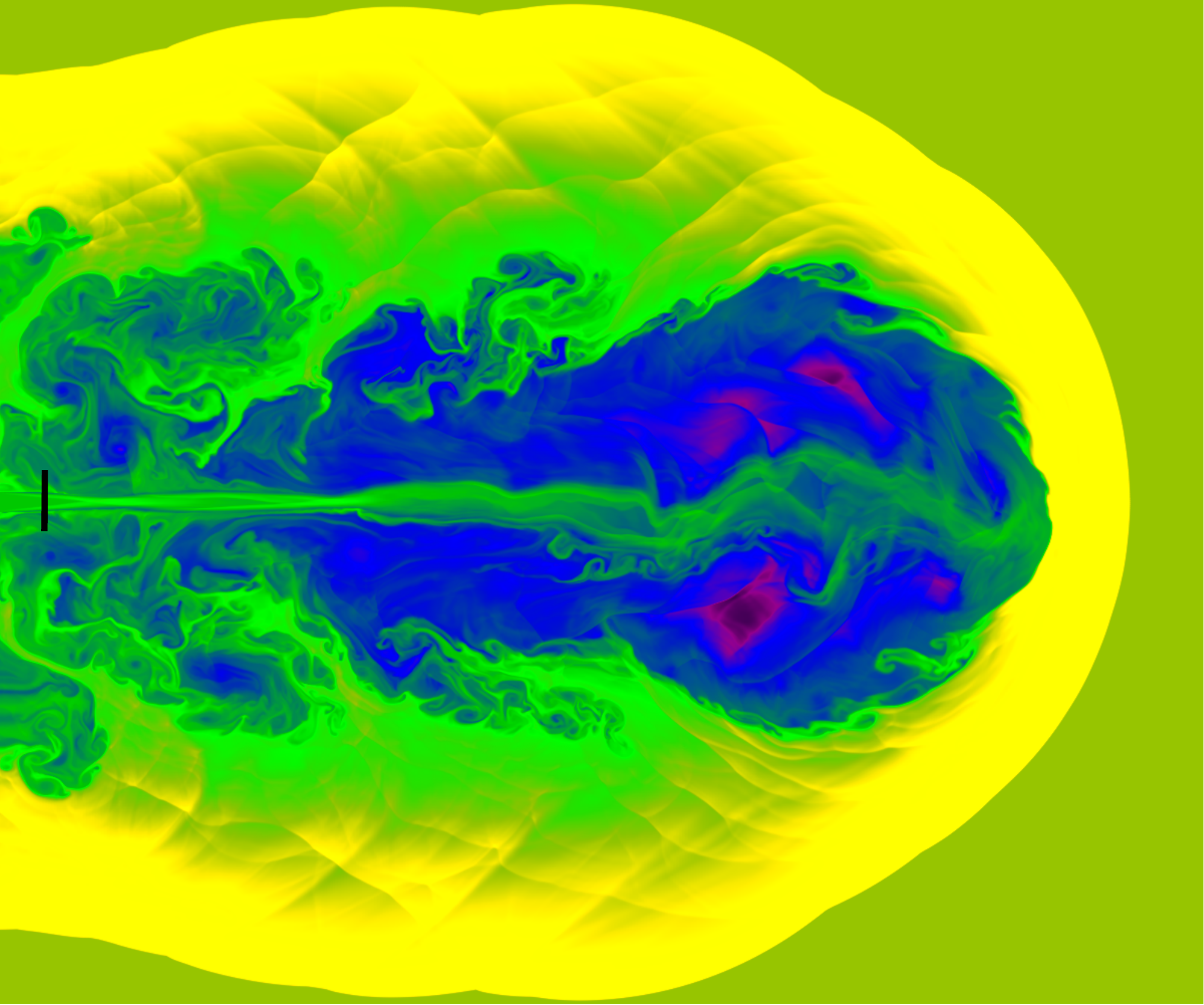, height=2.4in} & \epsfig{file=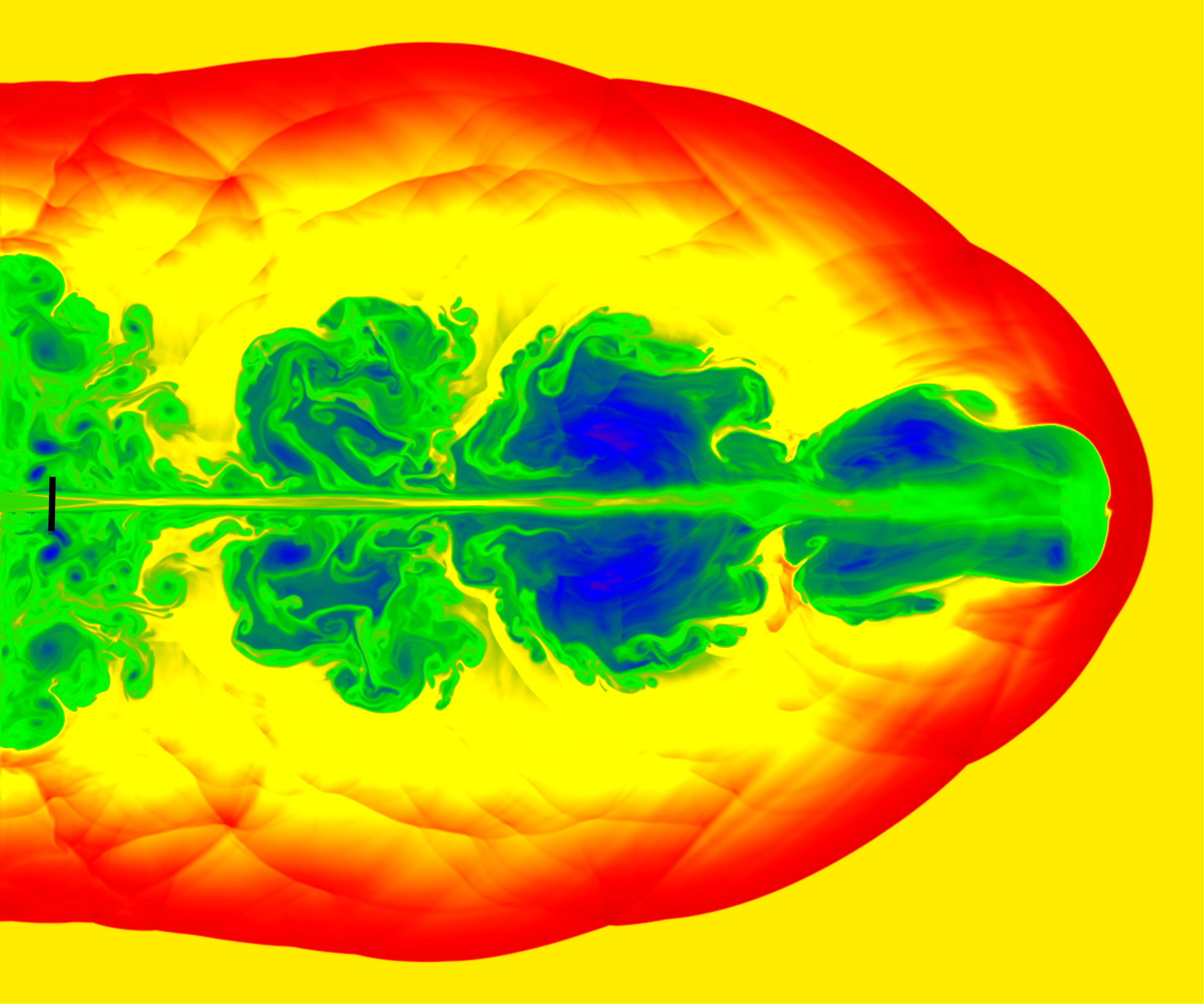, height=  2.4 in} \\	
	
																\end{tabular}	
										\caption{
										Snapshots of  fluid densities (on a logarithmic scale) for $v_j = 0.95c$ with density ratios, $\eta = 0.1$ (top left) and $\eta = 0.01$ (top right); $v_j = 0 .99c$ with density ratios $\eta = 0.01$ (middle left) and $\eta = 0.00316$ (middle right); 											$\eta = 0.01$, with $v_j = 0.995c$ (bottom left), and $v_j = 0.998c$ (bottom right).  For the first five images the scale is such that the maximum density (yellow) is $\simeq 10$ times that of the ambient density (green) and the minimum density (violet) is $\simeq 10^{-3.2}$ of the ambient density; for the final image the ambient density is yellow and  red corresponds to 10 times the ambient while violet is again $\simeq 10^{-3.2}$ of the ambient density. }  
						\end{center}
				\end{figure}

				After substantial experimentation with different parameters we found that our best overall results came from  simulations of 2D slab-like RHD jets on a $1200 \times 1000$ zone grid, with a jet width of 20 zones. We found that this resolution and jet width were sufficient to accurately model the blazar variability and to resolve shocks within the jet while still staying within the limits of our computational resources.   These jets differ from true 3D jets in that the jet power is spread over a thin vertical ``slab," rather than a circle of radius $r_{jet}$ (e.g.\ Hardee \& Norman 1988, 1990; Zhao et al.\ 1992).  This leads to an overall faster propagation of the jet for any given jet-to-ambient density ratio and jet velocity, as well as the obvious lack of development of 3D instabilities.   In these RHD simulations, the speed of light is the ratio of  1 distance unit to 1 time unit; hence if we adopt a jet width of 40 light years (lt-yr) with a resolution of 20 zones across the  jet, and take 2 zones to be equal to one distance unit, then one time unit is equal to 4 yr in the rest frame of the jet.  However, the physical time unit for these calculations scales directly with the chosen jet width.
				
				Hardee \& Norman (1988, 1990) discuss the spatial stability of the slab jet and compare the propagation of the slab jet analytically and numerically to a 3D cylindrical jet. Analytically, they found that the slab jet differed qualitatively from the cylindrical jet only by the harmonics of the helical modes, which do not have a counterpart in the slab jet. They found that the quantitative differences between the slab and cylindrical jets predominantly  result from the different  propagation angles of the perturbation with respect to the jet that are allowed by these different geometries.  When performing numerical simulations with slab and cylindrical jets, Hardee \& Norman (1988, 1990) found that overall the slab jet was a good analog to the cylindrical jet; however, certain modes in each type of jet contributed more or less than expected from the analytical results (also see Bassett \& Woodward 1995). Based on this work, we believe that for our purposes the slab jet  simulations we conduct provide sufficiently good  approximations to the growth of instabilities in 3D relativistic jets, though of course this is a crucial simplification that deserves further investigation.  
				
				Most earlier simulations of propagating jets employed grids with lengths (along the jet direction) substantially greater than their widths (normal to it) so as to see a more substantial portion of the jet (e.g.\ Wiita \& Norman 1992; Hooda et al.\ 1994; Carvalho \& O'Dea 2002a,b).  However, we found our RHD runs reached a far greater number of timesteps without terminating when the grid dimensions were about equal in width and length. It is possible that this is because less of the matter driven by the bow shock is lost off the sides of the grid, but this necessarily reduces the ratio of the length of the simulation to only 60 jet diameters. 
				
				Each simulation was run at least until any of the following conditions were satisfied: (i) the bow shock reached the far end of the grid;   (ii) the jet clearly lost stability;  (iii) the Athena code terminated before either of those more desired outcomes.  We chose to run simulations for a range of jet-to-ambient density ratios ($\eta$) from 0.001 to 0.1 and a range of initial jet velocities ($v_{jet}$) from $0.70c$ to $0.999c$.  These inputs for our models are summarized in Table 1 along with the corresponding values of $\gamma_j$, $\eta_c$, and $M_j$, as defined in Eqns.\ (1) and (2) below. Simulations that maintained stability for long enough to provide sufficient resolution in the final power spectra derived from the light curves are noted.  Snapshots of six of these runs are given in Fig.\ 1 where the logarithms of the gas density are displayed when the jets have progressed most of the way across the grid.  Unfortunately, our limited computational resources precluded us from using the much larger grid that would have allowed us to examine variations after a quasi-steady state jet had developed for the parameter sets that remained stable while the propagating jet completely crossed the grid.  We also did not continue the simulations for times much after the leading edge of the jet left the grid because the loss of material from the grid clearly affects the outcome of the simulations.  To some extent this limits the validity of our models because they are affected somewhat by initial transients in the jet propagation. However, because the region we analyze is behind the quasi-stable reconfinement shock, which is not very far along the jet, and we only start computing light curves after that shock is established and only compute power spectra while the jet appears to remain stable, we do not believe any transient aspect of the simulation would make much of a difference in the results we present below.

	\begin{deluxetable}{ccccc}
	\tablenum{1}
	\tablecaption{Simulation Parameters}
	\tablewidth{0pt}
	\tablehead{				
	\colhead{$\beta_j$} & 	\colhead{$\gamma_j$} & \colhead{$\eta$} & \colhead{$\eta_c$} &  \colhead{ $M_c$ }\\
	}
	\startdata
						0.70 & 1.40 & 0.01 & 0.0217 & 5.507  \\
						\textbf{0.90} & \textbf{2.29} & \textbf{0.1} & \textbf{0.5816} & \textbf{11.60} \\ 
						\textbf{0.90} & \textbf{2.29} & \textbf{0.01} & \textbf{0.0582} & \textbf{11.60}  \\ 
						\textbf{0.95} & \textbf{3.20} & \textbf{0.1} & \textbf{1.1333} & \textbf{17.10}  \\
						\textbf{0.95} & \textbf{3.20} & \textbf{0.01} & \textbf{0.1133} & \textbf{17.10}  \\
						0.95 & 3.20 & 0.00316 & 0.0358 & 17.10  \\
						0.99 & 7.09 & 0.1 & 5.553 & 39.43  \\
						\textbf{0.99} & \textbf{7.09} & \textbf{0.01} & \textbf{0.5553} & \textbf{39.43}  \\
						\textbf{0.99} & \textbf{7.09} & \textbf{0.00316} & \textbf{0.1755} & \textbf{39.43}  \\
						0.99 & 7.09 & 0.001 & 0.0555 & 39.43 \\
						0.995 & 10.01 & 0.1 & 11.08 & 55.98 \\
						\textbf{0.995} & \textbf{10.01} & \textbf{0.01} & \textbf{1.108} & \textbf{55.98}  \\
						0.995 & 10.01 & 0.00316 & 0.3501 & 55.98  \\
						0.995 & 10.01 & 0.001 & 0.1108 & 55.98 \\
						\textbf{0.998} & \textbf{15.82} & \textbf{0.01} & \textbf{2.765} & \textbf{88.71}  \\
						0.998 & 15.82& 0.00316 & 0.8745 & 88.71 \\
						0.998 & 15.82 & 0.001 & 0.2765 & 88.71  \\
						\textbf{0.999} & \textbf{22.37} & \textbf{0.01} & \textbf{5.528} & \textbf{125.6} \\
						0.999 & 22.37 & 0.00316 & 1.748  & 125.6  \\
	\enddata
	\tablecomments{Simulations that were stable and ran long enough to be deemed useful for Power Spectral Density analysis are bolded.}
	\end{deluxetable}

				In Table 1, $\eta_c$ and $M_c$ are the equivalent classical density ratios and Mach numbers, respectively (Rosen et al.\ 1999).  In the formulation of  Carvalho \& O'Dea (2002a), 
\begin{equation}			
				 \eta_c = \gamma_j^2 \eta (1 + \delta_2), 
\end{equation}
\begin{equation}	
				M_c = \sqrt{3} \gamma_j \beta_j (1 + \delta_2^{-1})^{1/2};
\end{equation}
\noindent here $\delta_2$ is a factor related to the energy density of relativistic electrons, and is defined as
\[				
				\delta_2 = \frac{4}{3}(1+k)\bar{\gamma}_e \frac{m_e}{m_p}q ,
\]	
					
				\noindent where $k\approx 0$ for a pair plasma jet (Case 1, Carvalho \& O'Dea 2002a), while $k \approx 100$ is typically taken for a proton-electron jet (Case 2), $\bar{\gamma}_e$ is the average Lorentz factor of the relativistic electrons, and $q$ is the ratio of the number of electrons to total particles within the jet, taken to be 0.5.  For our calculations, a $\delta_2$ value of 0.105 was used, in approximate accordance with their Case 1. 
We took the jet diameter to be 40 lt-yr, characteristic of an early stage radio jet that could be resolved by VLBI.  This also produces a plausible location from which jet variability might arise over timescales on the order of months to many years in the observer's frame.

			\section{Modeling Variability}
	
				\subsection{Bulk Velocity Variability}
				A substantial portion of the long-timescale variations in the light curves of blazars can be attributed to slight changes in the direction of the fluid flow with respect to the observer (e.g.\ Camenzind \& Krockenberger 1992; Gopal-Krishna \& Wiita 1992). Any radiating material traveling at relativistic speed will experience Doppler boosting, which strongly beams the emission in the direction of its velocity.  As the flow in the jet wiggles back and forth, the angle of the velocity with respect to the observer changes, resulting in the apparent rise and fall of the received flux.  The Doppler boosting factor is given by
\[
				D = \frac{1}{\gamma (1-\beta \cos \theta)}, 
\]
where $\gamma$ is the bulk Lorentz factor, $\beta$ is the speed in units of $c$, and $\theta$ is the angle of the velocity with respect to the observer. The observed flux is proportional to the $s + m$ power of the Doppler boosting factor, where $s$ is the slope of the synchrotron spectrum (with $S_{\nu} \propto \nu^{-s}$), and $m$ is 2.0 for continuous flow and 3.0 for shocked flow (e.g. Begelman et al.\ 1984).   We took $s$ to be equal to 0.5 for the jet, and $m$ was set to 2.0.  The fluid was assumed to be emitting isotropically  in its rest frame, but the changing Doppler boosting produces varying degrees of amplification in the observer's frame.  In the absence of explicit magnetic field strengths and directions in our models, the assumption of tangled fields and thus rest-frame isotropy for the emission is the default (e.g.\ Mart{\'i} et al.\ 1997; Mioduszewski et al.\ 1997).
				
				To draw data for use in this variability analysis, we chose a ``slice," 1 zone thick, spanning the width of the jet. The horizontal location of this slice was fixed and chosen to lie at a point just behind  (upsteam of) the first, and nearly steady, recollimation shock.  This shock typically forms very soon after the jet goes beyond $\sim 100$ zones or $\sim 5$ initial jet diameters.    The locations of the slices were determined visually after examining the entire simulation for the location of the internal (reconfinement) shock and are shown in Fig.\ 1.  While the shock location does vary somewhat, it is relatively stable, and so we picked a slice that always remained upstream of the center of the shock.  Since the plasma is being compressed in this region the emissivity goes up, not merely through the presence of higher densities of relativistic particles and stronger magnetic fields (though, recall, we do not track the latter), but also because the relativistic electrons are being accelerated to higher  energies and hence the flux from this region is higher in comparison to most other regions of the jet.  We determined which zones were within the jet by setting a threshold for the horizontal component of the velocity equal to $0.8  v_{j}$, and cells that did not exceed this forward speed were excluded from our calculations.   This choice is somewhat arbitrary but we made it to restrict the emissivity calculations to zones that are comprised predominantly of the jet's relativistic plasma and not those contaminated and slowed by external gas entrained by the jet.
				
				For this  portion of the variability modeling depending on jet direction  and speed we used a range of output timesteps starting when the recollimation shock had reached stability and ending either once the simulation terminated or the jet lost stability.  In total, this was typically on the order of 1500-3000 output timesteps, which was sufficient for use in the calculation of power spectra.  For each zone within each timestep, we drew from its velocity data and calculated a Doppler boosting factor.  This factor was then raised to the 2.5 power ($s + m$) and multiplied by the emitted flux.  This basic flux per unit volume  can be considered arbitrary for our purposes, so it was set to unity for an individual zone and then scaled by the zone density taken to be indicative of the number of relativistic electrons available to radiate.  The resulting observed flux was then summed along with the observed fluxes of the other zones within the slice from that timestep to yield a total observed flux for that time and then normalized by dividing by the number of zones so the plotted mean observed flux is unity.   Results of several of these light curve simulations are shown in Fig.\ 2 for a viewing angle of $\theta = 5^{\circ}$.
Taking into account this small angle and the high jet bulk velocities for all our simulations, in the observer's frame the light travel time from one side of the jet to the other is less than   the length of an output timestep in the simulation and therefore any time delay effects can be ignored for these variations.    By only computing the emission from a small region of the jet (as was also done by Marscher 2014) we are ignoring the sum of the radiation produced by the remainder of the jet, which will typically substantially exceed that from the region we can analyze but is not expected to vary as much.  Hence, the actual observed variability will be damped by the less variable emission from those other regions and in that sense the changes in the plotted light curves are  exaggerated.  However, because power spectra are computed after the mean flux is subtracted, their shapes and slopes are not affected by an overall increase in the baseline emission.

				\begin{figure}
					\begin{center}
						\begin{tabular}{c c}
							\epsfig{file=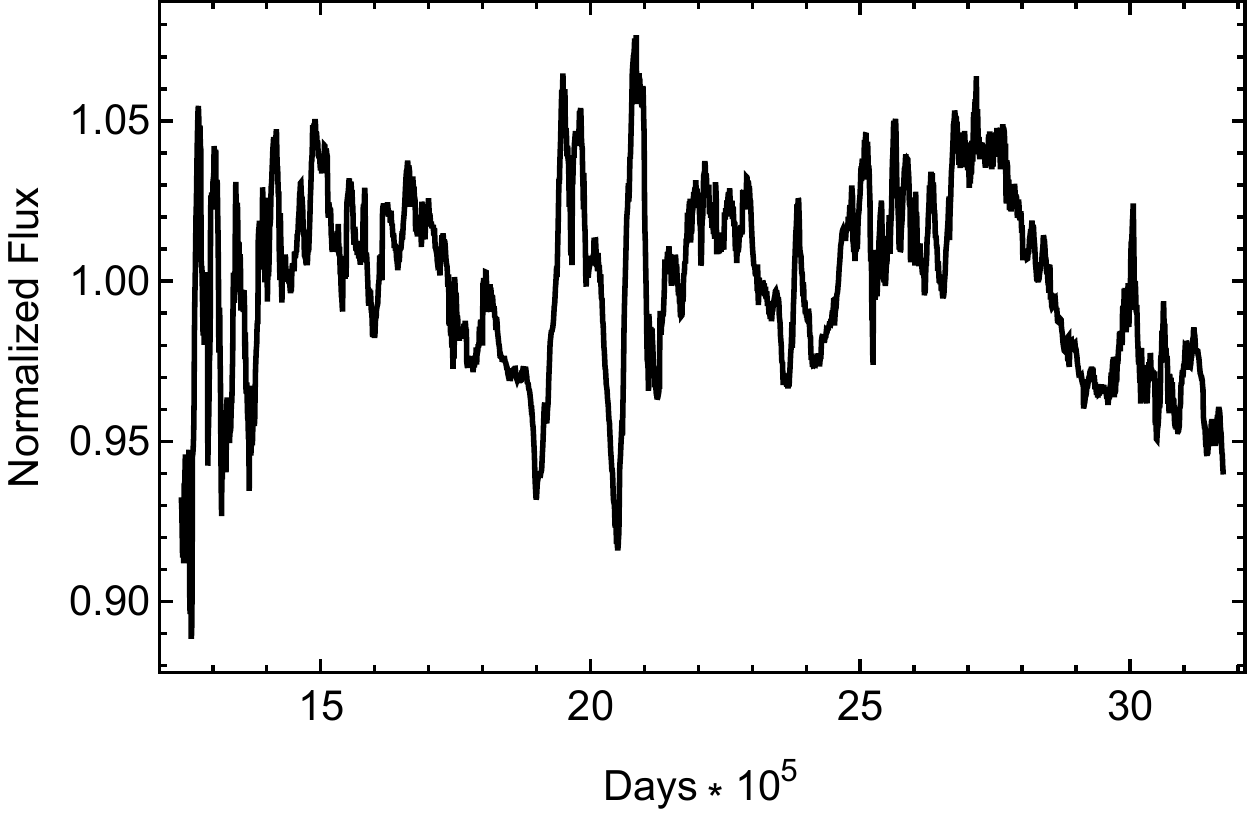, height=1.5in} & \epsfig{file=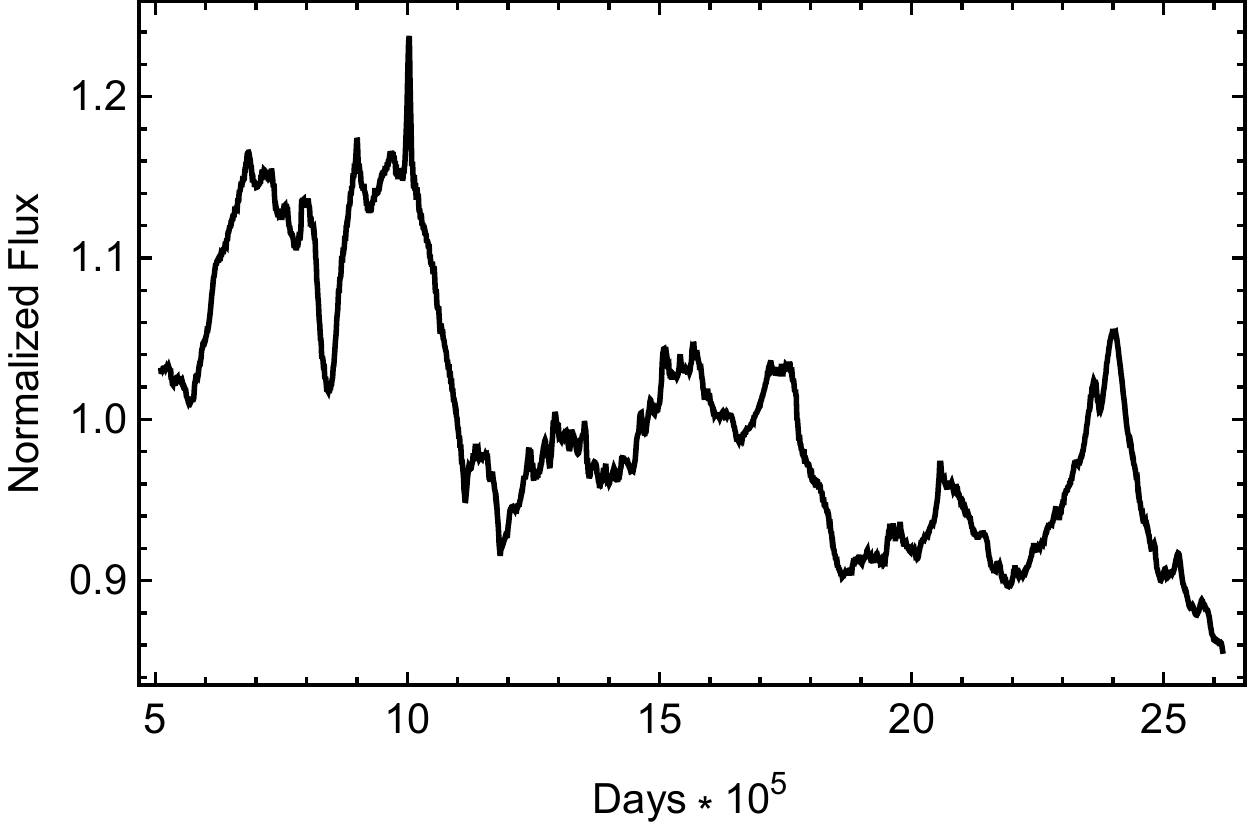, height=1.5in} \\
							\epsfig{file=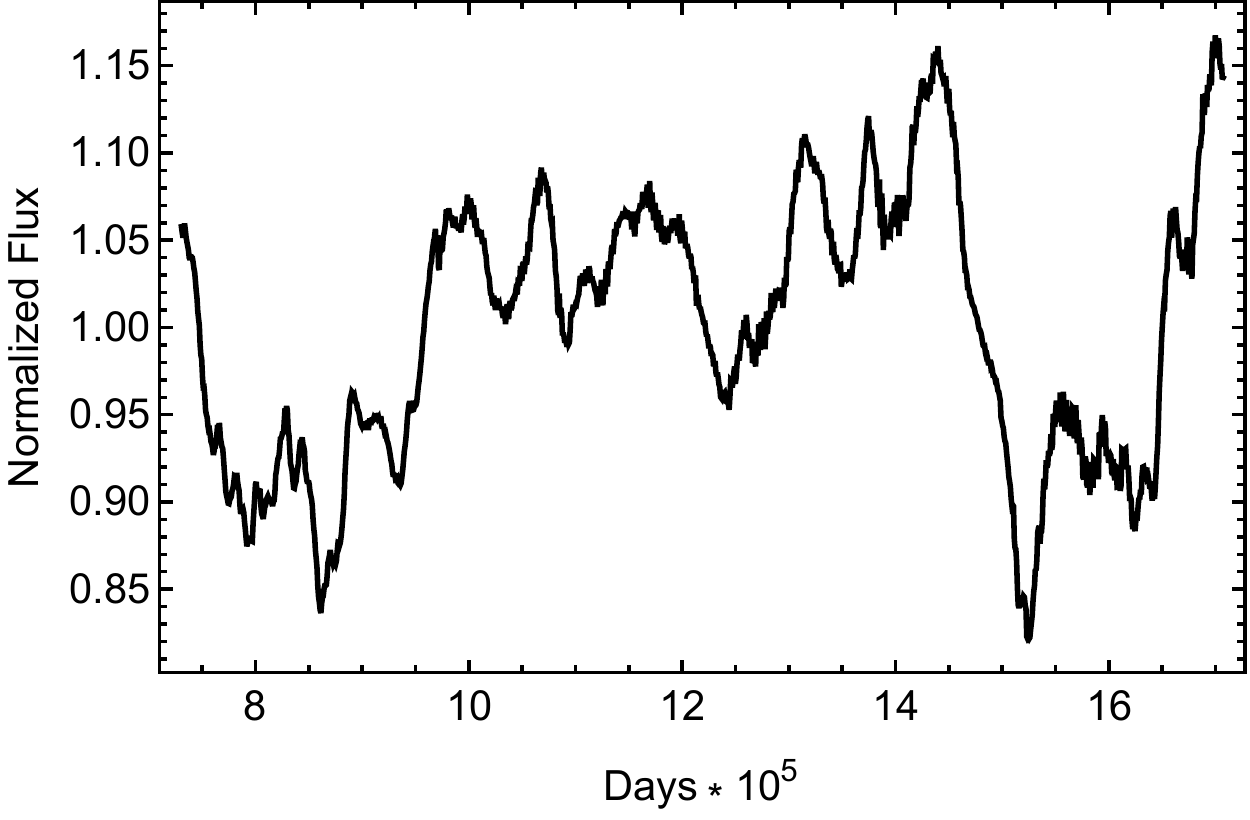, height=1.5in} & \epsfig{file=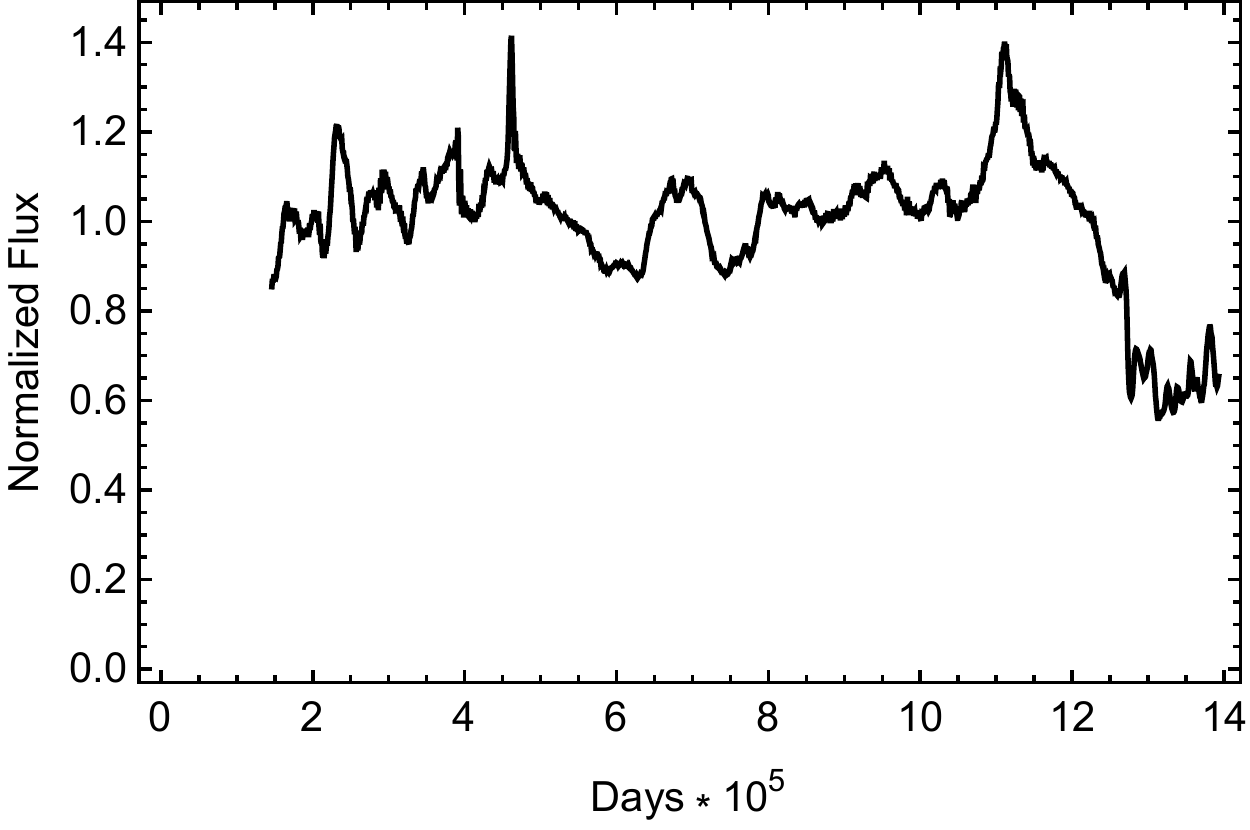, height=1.5in}
					
						\end{tabular}
						\caption{Bulk velocity light curves for jets with (top left) $v_j = 0.9c$ and $\eta = 0.01$; (top right)  $v_j = 0.95c$ and $\eta = 0.1$; (bottom left) $v_j = 0.99c$ and $\eta = 0.00316$; (bottom right) $v_j = 0.995c$ and $\eta = 0.01$.}
					\end{center}
				\end{figure}

				\subsection{Turbulent Variability}
				
				On a  timescale substantially shorter than that arising from jet wiggles, there are significant variations in the observed flux due to turbulence within the flow of the jet. As noted above, in our fluid-dynamical simulations this turblence is sub-scale and so not included in the dynamics. Hence we need to model independently the effects of turbulence on variations in the light curve and then see whether or not the fluctuations produced by the turbulence have similar power spectra to those produced by the bulk velocity variations and if they agree with observations.  We took the turbulence to be comprised of successively smaller levels of refinement of eddies;  each  such smaller portion of fluid spins around at some velocity that scales with eddy size, and is characterized by a maximum turbulent velocity, $ v_t $ corresponding to the largest eddy. This spinning fluid is constantly changing direction and therefore also changing its Doppler boosting factor, which results in variations in the observed flux coming from the jet on the order of days, which is roughly the turnover time of the smaller eddies.  
				
				The turbulence variability code used in CW neatly accomplishes this calculation by taking an input bulk velocity $v_b$, maximum turbulent velocity $v_t$, and observation angle $\theta$ and computing light curves for nine levels of eddy refinement; the largest eddy size was taken to correspond to the width of one jet zone. Their code incorporates the time delays arising from the difference in light travel time to the observer from eddies on one side of the cell to the other.  Additionally, the code calculates the spectrum of turbulent velocities relativistically, instead of from the classical Kolmogorov spectrum.  The Lorentz factors ($\gamma_t$) of the turbulent cells are modeled following Zrake \& MacFadyen (2013) as
\begin{equation}				
				 \frac{(\gamma_t-1)^3(\gamma_t+1)}{\gamma_t^2} = \frac{C \epsilon^2 l^2}{c^6} .
\end{equation}
Here, $C$ is a constant, $\epsilon$ is the rate of energy injection per unit mass, and $l$ is the size of the eddy.  The combination $C \epsilon / c^6$ is calculated from the provided maximum turbulent velocity $v_{t}$ by inserting its Lorentz factor (computed in the rest frame of the overall cell) into Eqn.\ (3) with $l$ as the size of the largest eddy. 
				
				We modified the CW code so as to incorporate fluxes emitted from multiple turbulent cells.  As input, the code was fed the velocity data from a single cell in the same ``slice" used in the bulk velocity variability calculations, except that only a single timestep from the simulation was used; it was always taken from a time at which the recollimation shock had reached stability. 
This was then repeated for each zone in the slice so that multiple light curves were produced. In that the timesteps modeled by the CW code were markedly shorter than those of our bulk velocity variability code, and more importantly, shorter than the light travel time from one side of the jet to the other, we did have to take time delays into account in the summation of these light curves.  This was accomplished by staggering the light curves, before the summation, by time intervals equal to the light travel delay between each zone, which was proportional to the sine of the assumed observation angle.  Several of the individual zone turbulent light curves are shown in Fig.\ 3; again we took  $\theta = 5^{\circ}$ for all displayed, as it is an appropriate value for blazars.
				\begin{figure}
				\begin{center}
					\begin{tabular}{c c}
				\epsfig{file=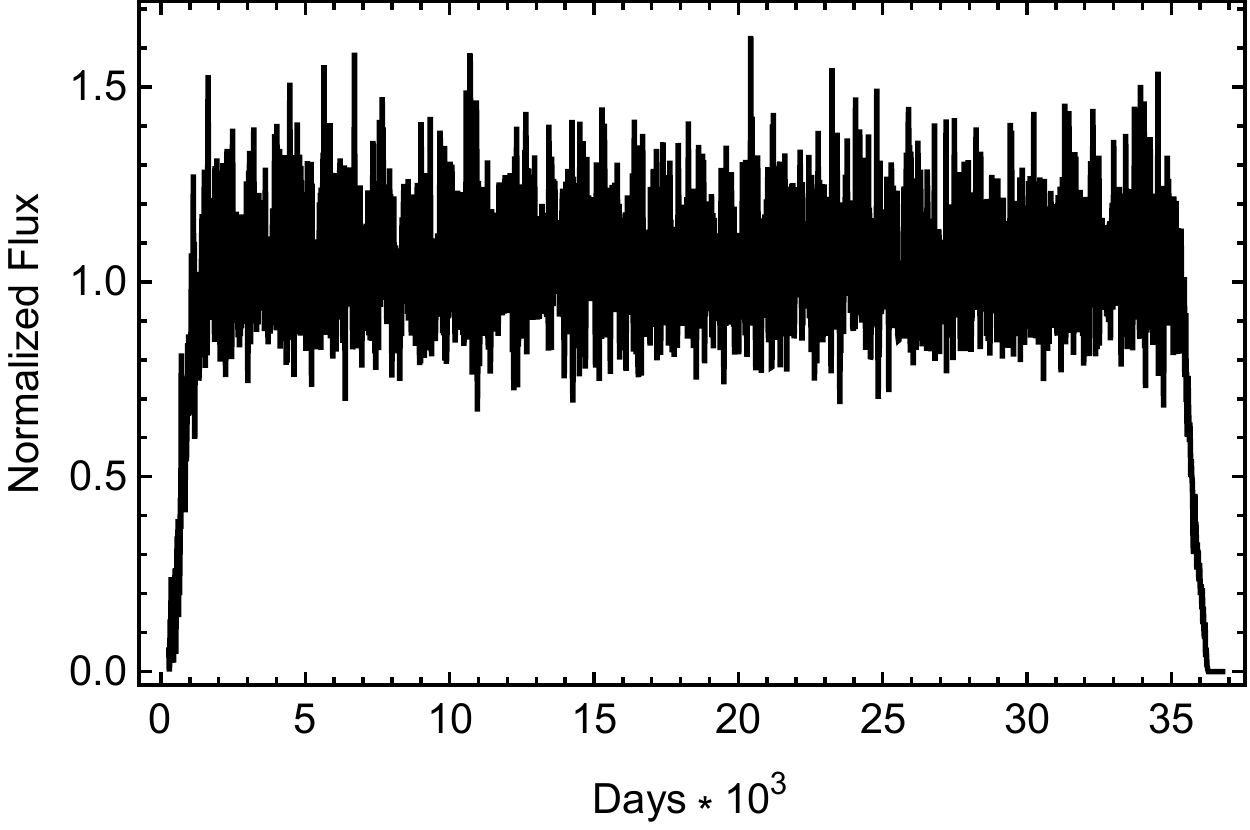 , height = 1.5in } & \epsfig{file=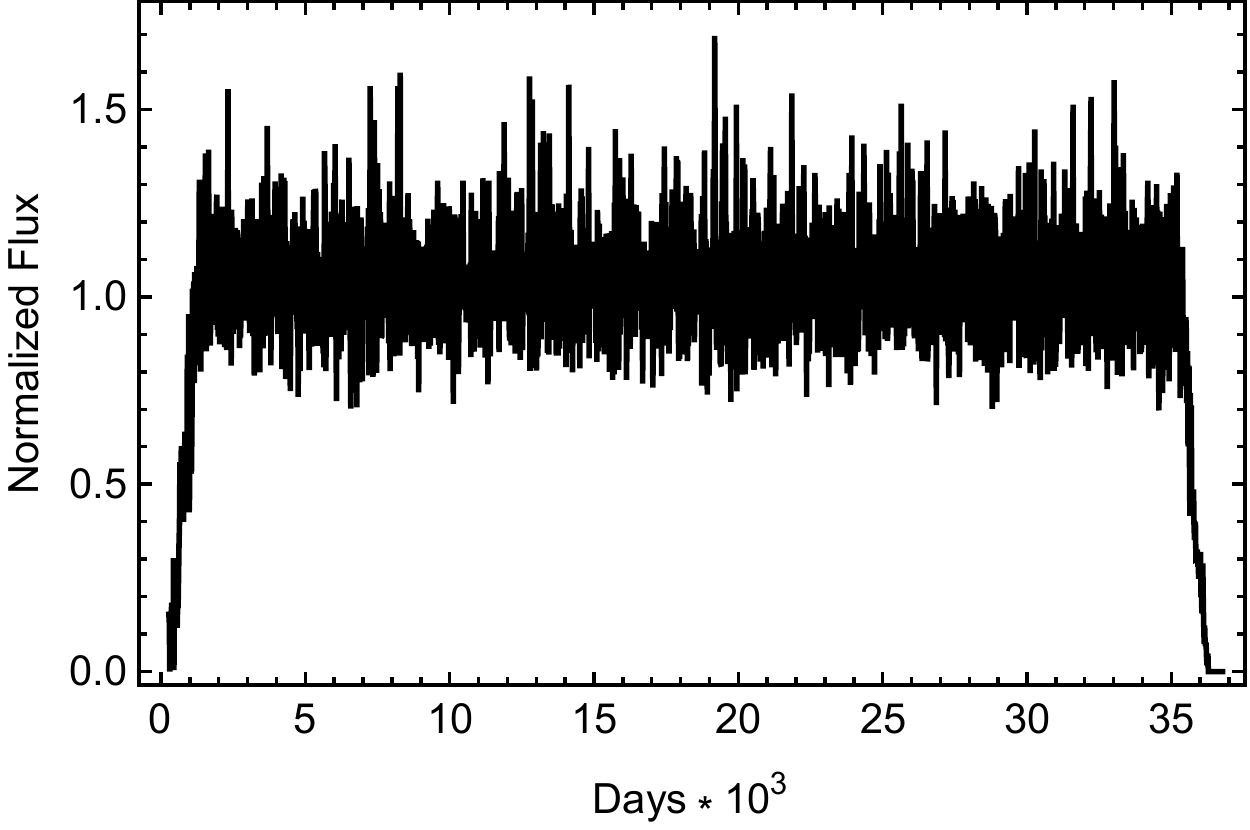 , height = 1.5in } \\
				\epsfig{file=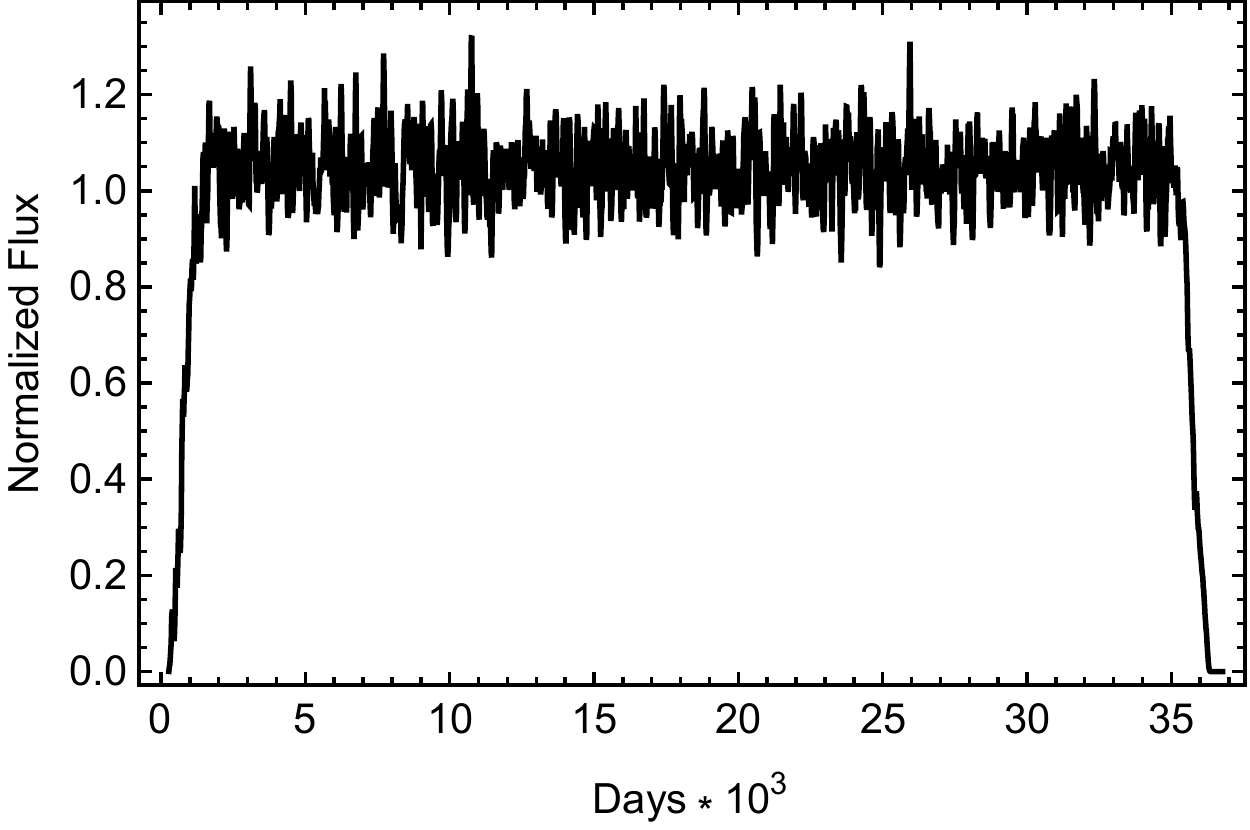 , height = 1.5in } & \epsfig{file=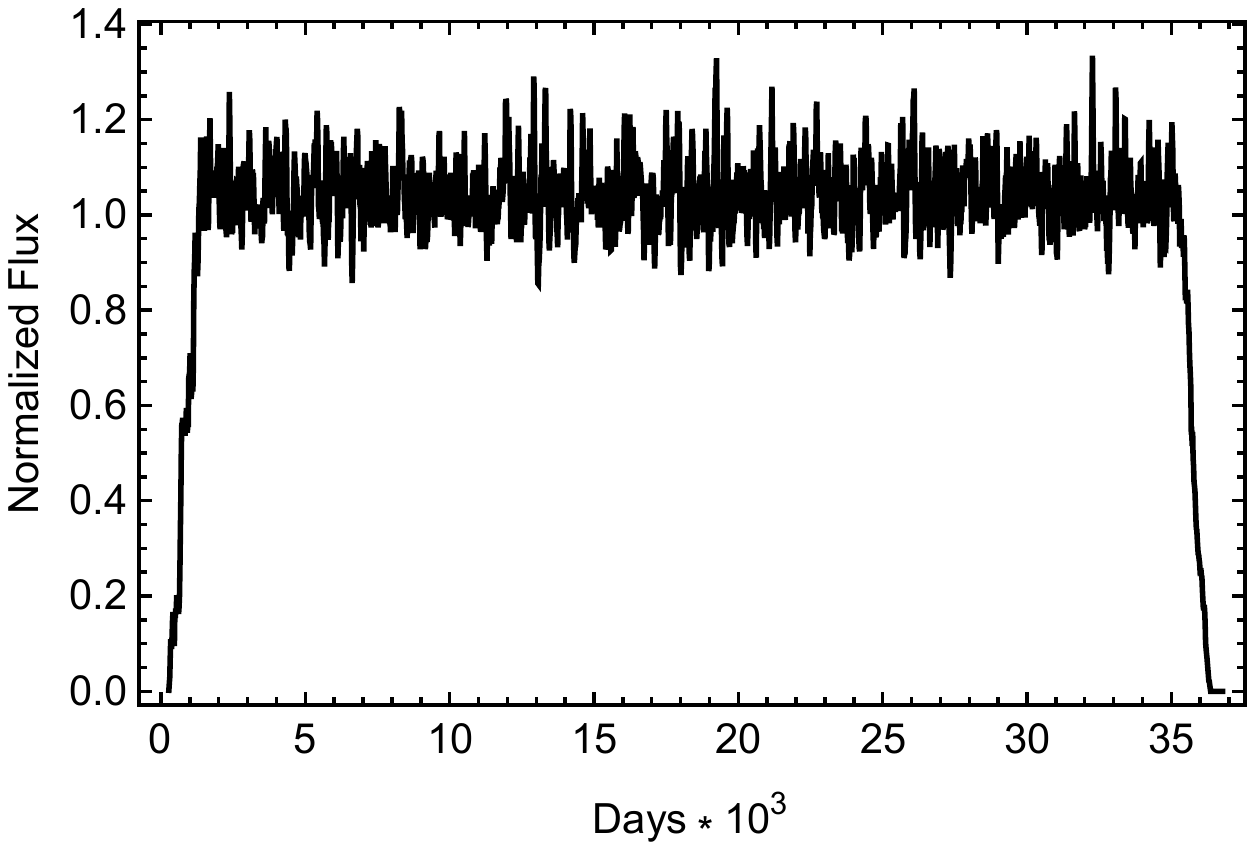 , height = 1.5in }
				\end{tabular}
				\caption{Turbulence light curve for a single cell of a jet with: (top left)  $v_j = 0.95c$ and $\eta = 0.1$; (top right) $v_j = 0.99c$ and $\eta = 0.00316$; (bottom panels) turbulence light curve summed over all cells in the slice with parameters as above.}
				\end{center}
		   		\end{figure}

				\subsection{Power Spectral Densities}
				
				For both of the light curves produced by the bulk velocity variability and the turbulent variability calculations, PSDs were computed.  First, the light curve data was normalized to the average value of the flux. Then we performed a Discrete Fourier Transform on the data to bring it over to the frequency domain. These values were squared to yield PSDs for the turbulent and bulk velocity analyses.  The PSDs of actual AGNs are usually found to be approximated by red noise at lower frequencies, with the power depending on frequency roughly as $P \propto f^{\alpha}$ with $\alpha \sim -2$, while at high frequencies, instrumental white noise ($\alpha \simeq 0$) dominates (e.g.\  Uttley \& McHardy 2005; Edelson et al.\ 2013; Wehrle et al.\ 2013; Revalski et al.\ 2014, and references therein).
				
				The nature of the Discrete Fourier Transform means that there were many fewer data points on the low end of the logarithmic frequency spectrum of the PSDs than on the high end. To mitigate this inequality when computing the slopes of the PSDs we binned the data points at five equal intervals per decade and computed an average for each bin, which we used to calculate the slope.  Beginning at lower frequencies, we checked each of these averages to find where the local slope exceeded a certain threshold value, after which all of the data points were excluded.  This was a fairly effective automatic method of excluding white noise and guaranteeing that Nyquist noise was eliminated from the PSD calculations, but did require some manual adjustment in certain cases.  The remaining averaged points for each bin were used to fit a power-law line to the data from the log-log plot of power against frequency to  calculate a final slope.  These slopes, determined separately for the bulk variability and turbulent variability portions of the computations, are given in Table 2 for the bulk simulations that were sufficiently lengthy to produce PSDs spanning over two decades.

\section{Results}

\subsection{Variability}

	\begin{deluxetable}{ccccc}
	\tablenum{2}
	\tablecaption{PSD Slopes}
	\tablewidth{0pt}
	\tablehead{				
	\colhead{$\beta_j$} & 	\colhead{$\gamma_j$} & \colhead{$\eta$} & \colhead{$\alpha_{turb}$} &  \colhead{$\alpha_{bulk}$} 
	}
	\startdata
							0.9 & 2.29 & 0.1 & -2.0 & -2.4   \\ 
							0.9 & 2.29 & 0.01 & -1.9 & -2.1 \\ 
							0.95 & 3.20 & 0.1 & -1.8 & -2.3 \\
							0.95 & 3.20 & 0.01 & -1.9 & -2.4 \\
							0.99 & 7.09 & 0.01 & -1.8 & -2.8 \\
							0.99 & 7.09 & 0.00316 & -1.8 & -2.9 \\
							0.995 & 10.01 & 0.01 & -1.9 & -2.6 \\
							0.998 & 15.82 & 0.01 & -2.2 & -2.2 \\
							0.999 & 22.37 & 0.01 & -2.3 & -2.3 \\
	\enddata
	\tablecomments{The slopes of the turbulent and  bulk velocity power spectra are $\alpha_{turb}$ and $\alpha_{bulk}$, respectively.}
	\end{deluxetable}

				\begin{figure}
				\figurenum{4}
				\begin{center}
				\begin{tabular}{c c}
				\epsfig{file=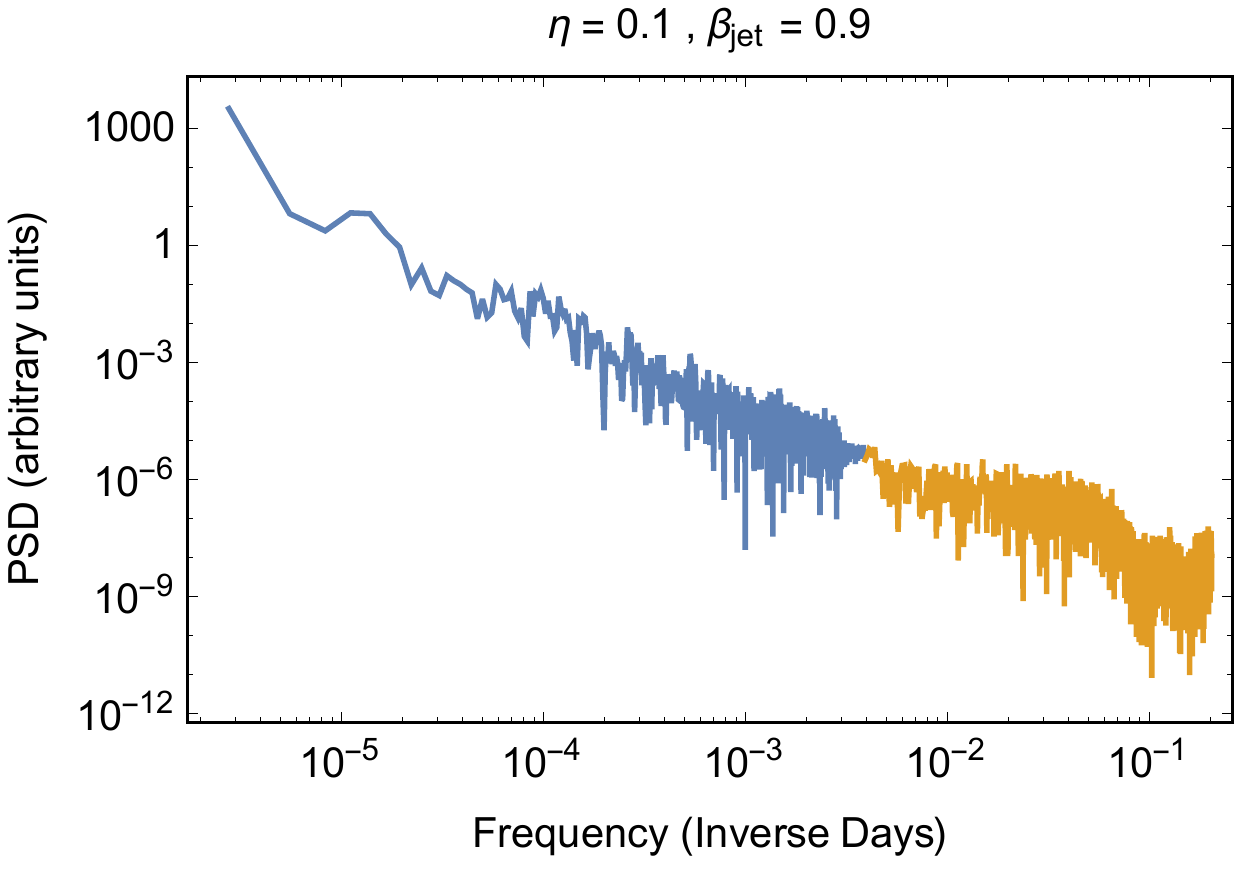 , height = 1.6in } & \epsfig{file=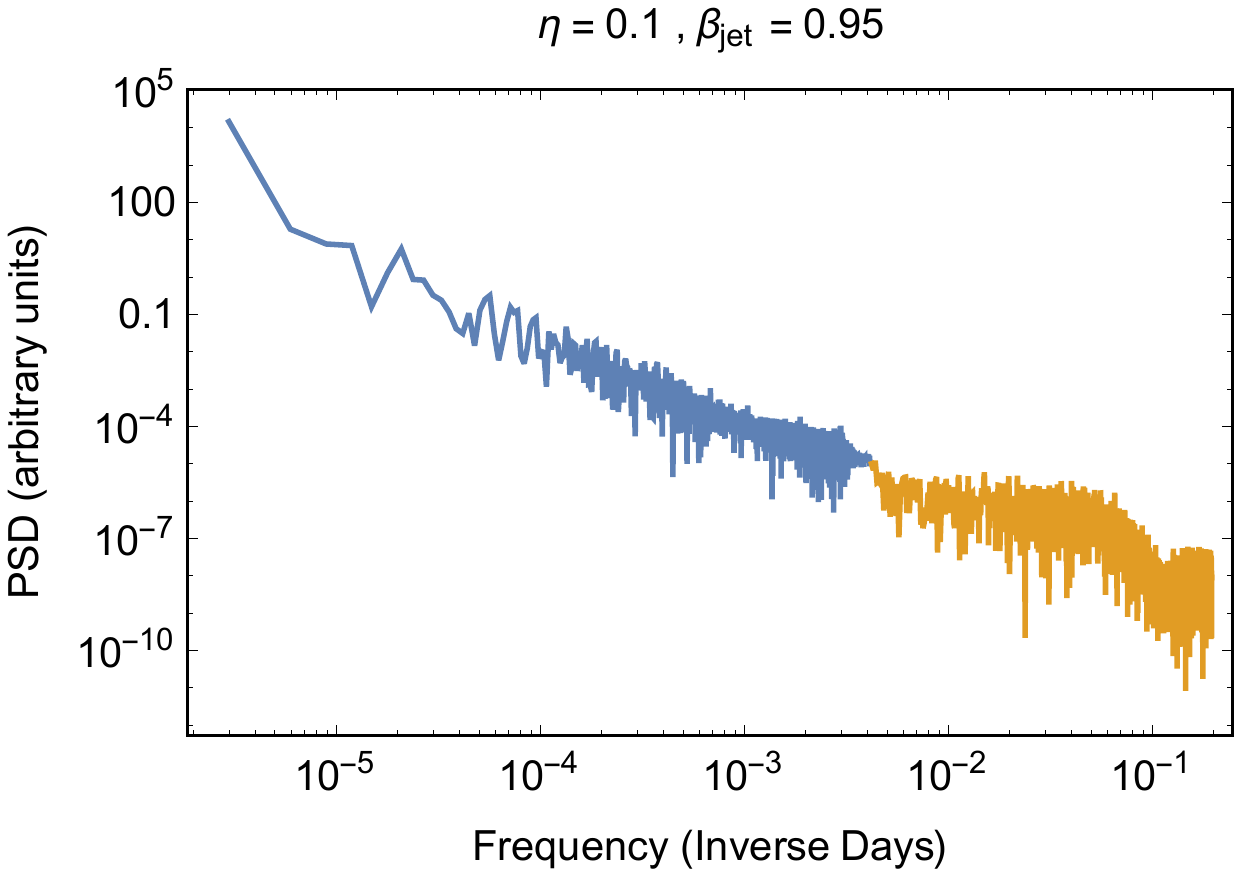 , height = 1.6in } \\
				\epsfig{file=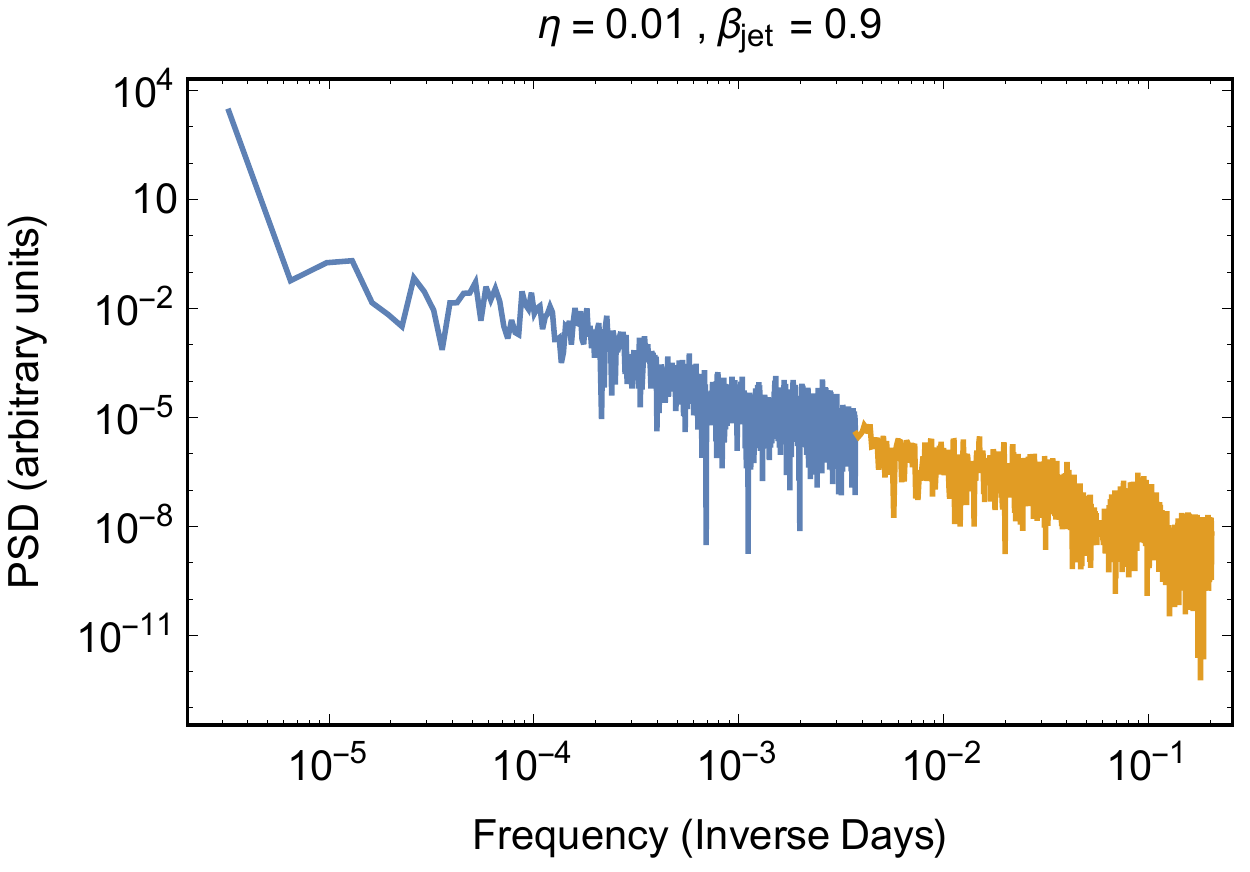 , height = 1.6in } & \epsfig{file=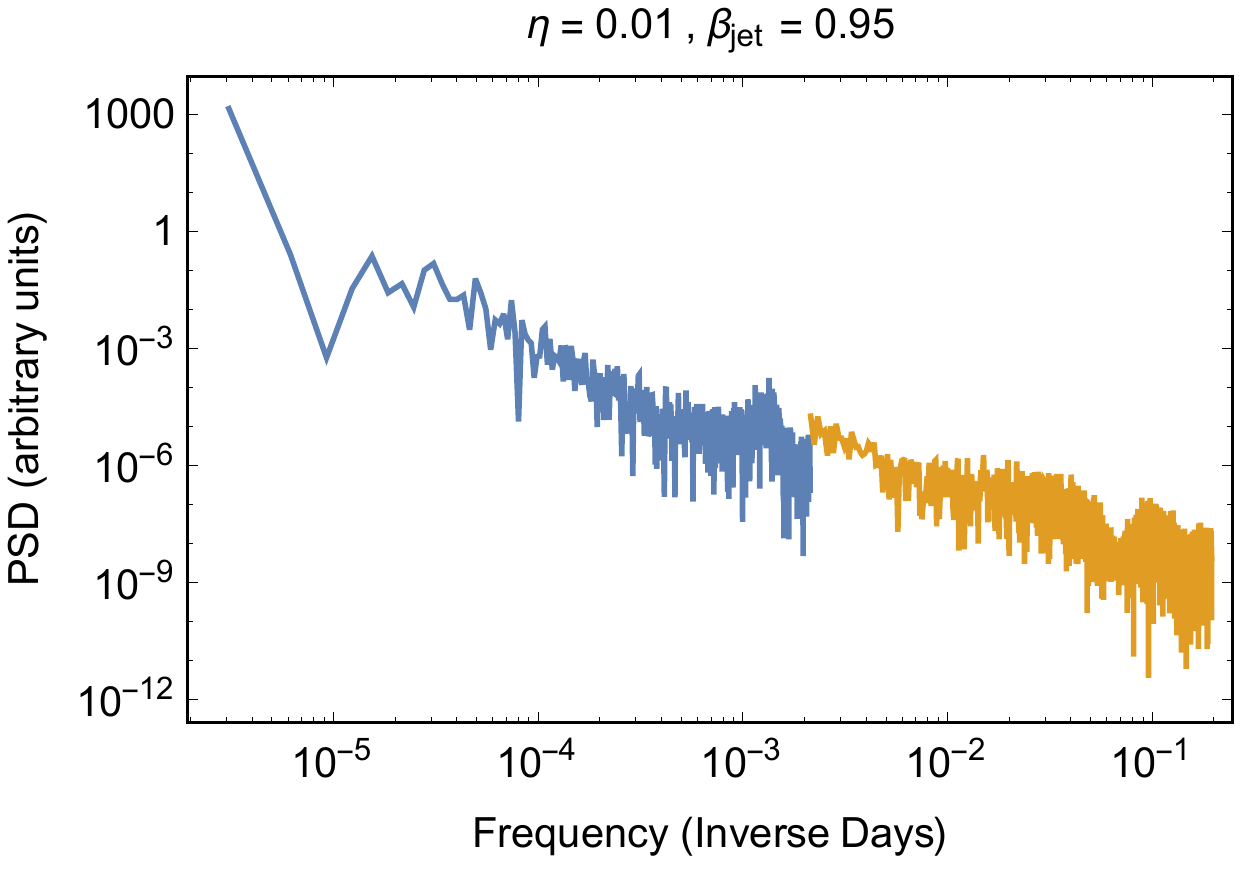 , height = 1.6in } \\
			    	\epsfig{file=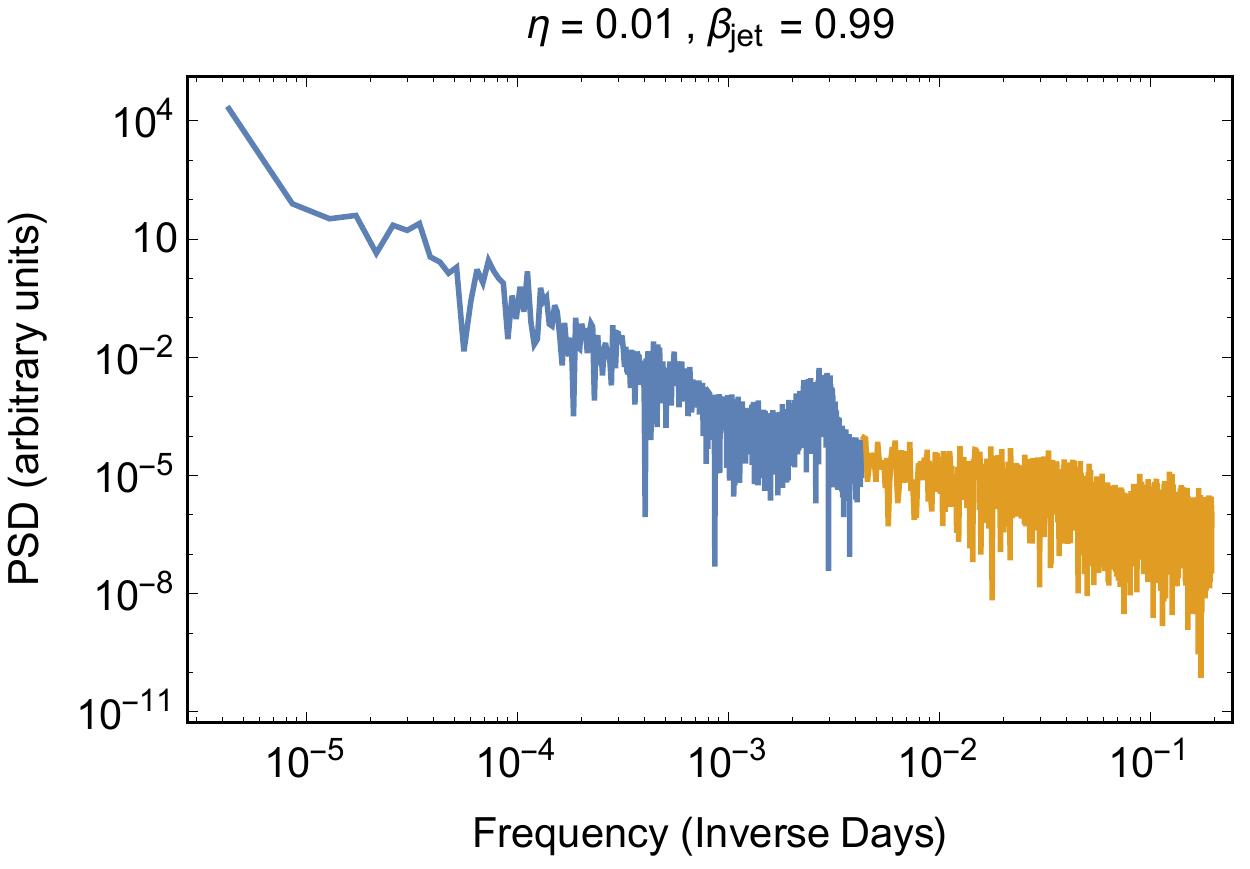 , height = 1.6in } & \epsfig{file=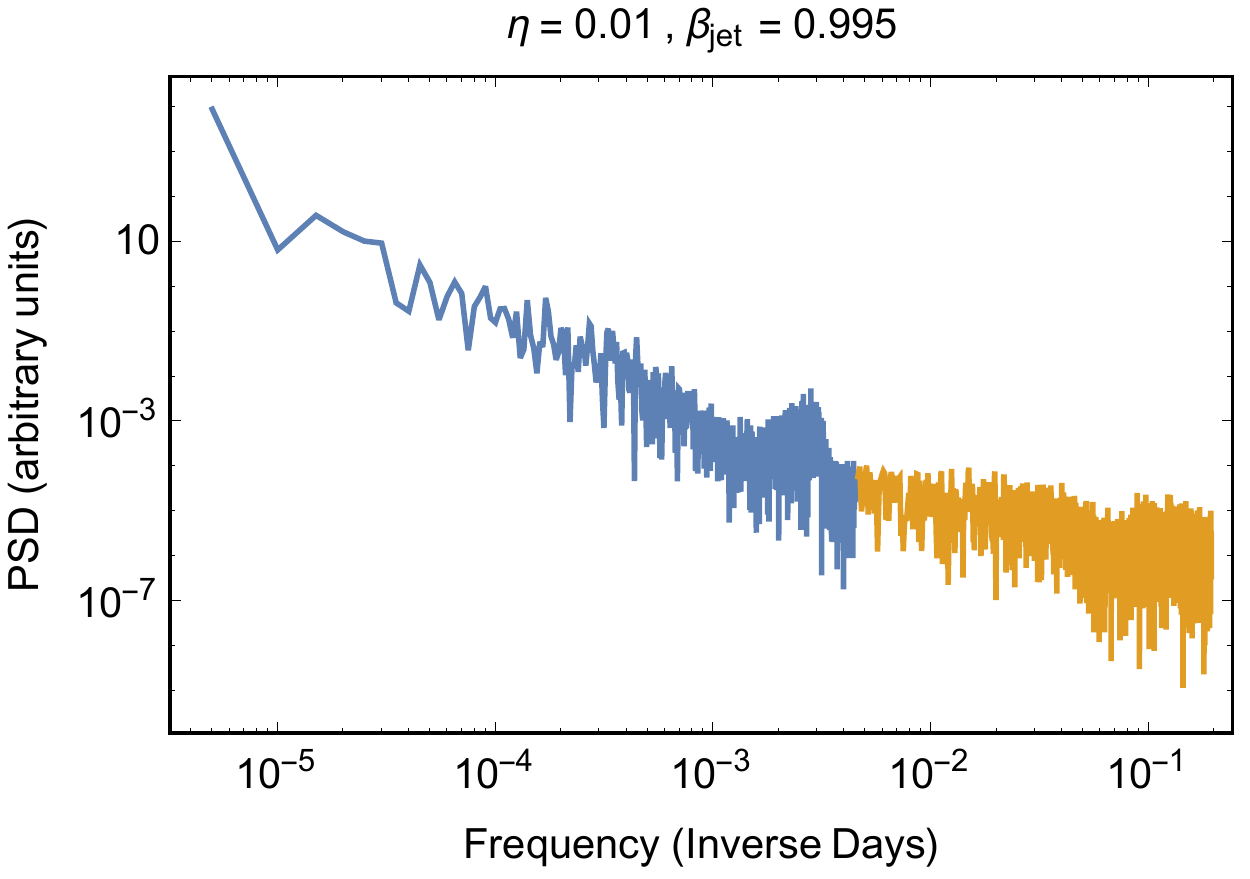 , height = 1.6in } \\
			    	\epsfig{file=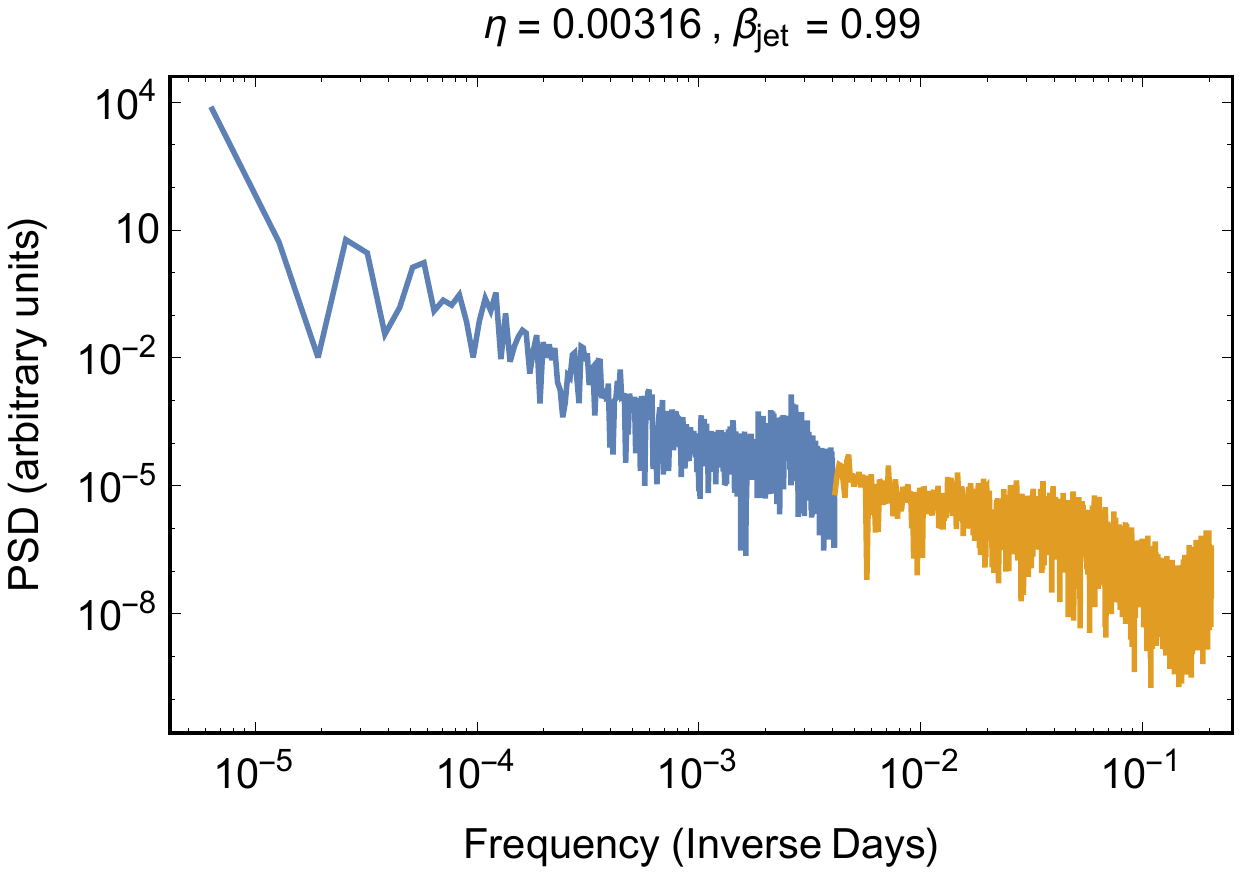 , height = 1.6in } & \epsfig{file=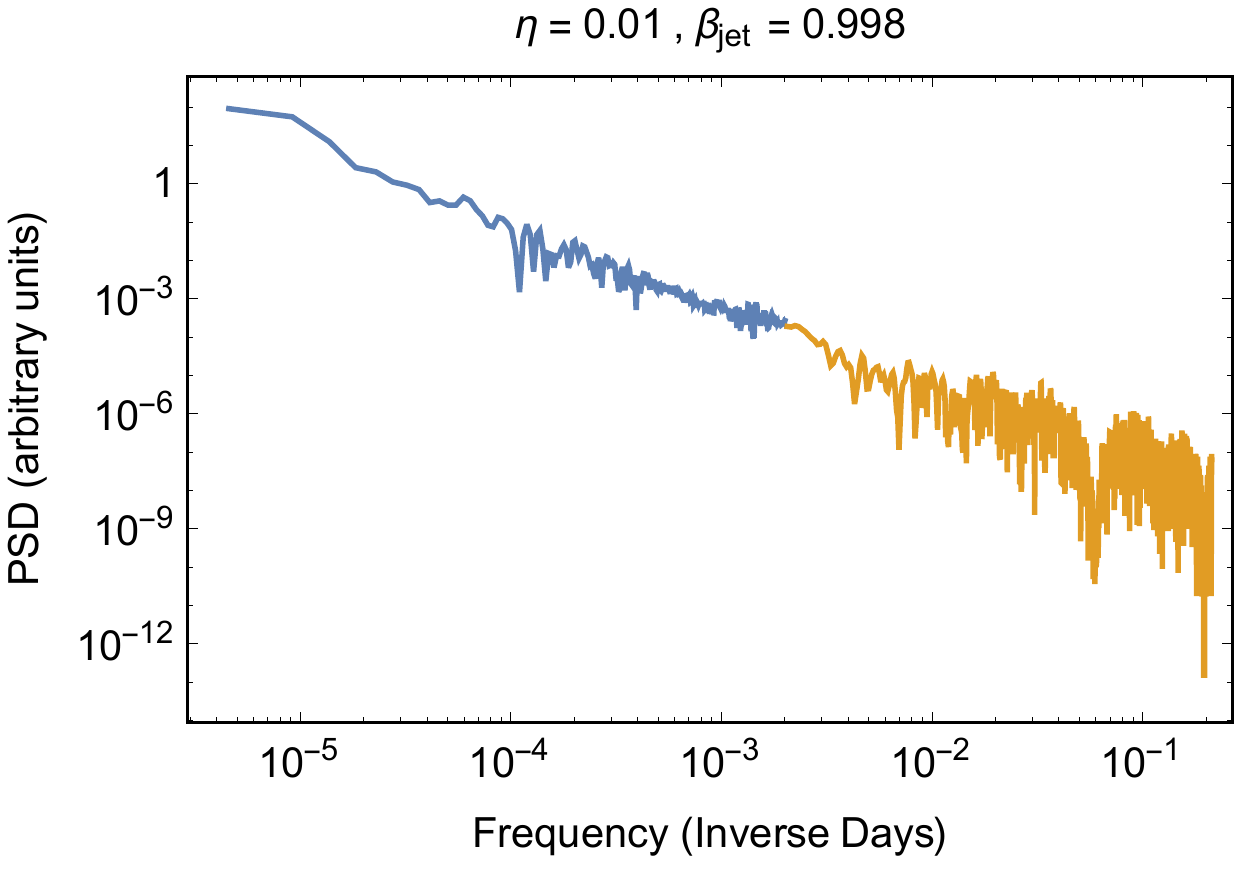 , height = 1.6in } 
				\end{tabular}
					\caption{The turbulent velocity (yellow) and bulk velocity (blue) PSDs, combined on the same plot, where the former is scaled to match the latter in the region of frequency overlap; values of $\eta$ and $\beta_j$ label each panel.}				
				\end{center}
				\end{figure}
				
				The light curves produced by the turbulence code did not vary greatly over any parameter with the exception of the maximum turbulent velocity $v_{t}$; this was in concordance with the findings of CW.  The value  $v_{t} = 0.3c$  yielded the most physically realistic light curves in that work and so was the one we adopted for all of the simulations treated here, although CW's simulations with lower maximum turbulent speeds also produced acceptable light curves.  Note that values of $v_t > c/\sqrt{3}$ would imply supersonic turbulence that is unlikely to be long-lived (e.g.\ Marscher 2014) and CW explicitly showed that values of $v_t > 0.6c$ produced light curves that displayed much larger positive excursions than negative ones and were unable to match observations.  The individual $v_t = 0.3c$ light curves varied around the average flux value by about 30\%, but as illustrated in in the bottom panels of Fig.\ 3, these variations are reduced when fluxes from many zones are averaged together.  Although we fixed the value of $v_t = 0.3c$ in these computations, in future work it would be preferable to allow for different values of $v_t$ to be taken from the bulk flow simulations, particularly in that the maximum turbulent velocity is unlikely to exceed the difference between the bulk velocities in adjacent zones or those of the same zone over the course of a few timesteps in their rest frames.   In our RHD simulations those differences in velocity between the grid zones in the slice are usually between $0.03c-0.2c$ and so the amplitudes of the turbulent variations displayed here are somewhat exaggerated.  Even though the turbulence is not clearly evident on the  scales  resolved by our bulk velocity grid, recent blazar observations justify the assumption that turbulence is present in the region upstream of the first confining shock (Marscher et al.\ 2008, 2010; Bhatta et al. 2013; Marscher 2014).  However, in light of our adopting the somewhat larger and fixed value of $v_t$, it is important to note that from the light curves and power spectra  considered in CW that while the amplitudes of fluctuations for $v_t = 0.1c$ are lower than for $v_t = 0.3c$, the slopes of the turbulent PSDs differ by no more than 0.1 between those two cases for otherwise identical parameters.  Therefore we are confident that our results would not be significantly affected by our use of $v_t = 0.3c$ instead of more precise values that might be obtained in future work from the differences in velocities between adjacent zones, particularly as the amplitudes of the light curves can be adjusted by changing the overall fraction of energy going into turbulence, which does not affect the slopes of the PSDs.

The sharp increases and decreases in fluxes at the starts and ends of the typical light curves in Fig.\ 3 are unphysical and correspond to the ``ramp-up" and ``ramp-down" times, respectively, that result from accounting for the light travel time delays.  When the turbulence simulation begins, it takes several timesteps for the radiation from the most distant eddies to reach the observer, and therefore the flux starts at zero and rises as photons from more and more eddies are received.   Once sufficient time has passed, light from all cells has reached the observer and the flux sensibly varies around a steady value.  At the end of the simulation all of the cells have stopped emitting, but light from the more distant ones still reaches the observer for varying times and so the flux does not cut off abruptly.   
To avoid anomalies induced by those artificial rises and falls,  the power spectra are computed from light curves with those portions of the fluxes excluded. As seen in Table 2, the slopes of the turbulent power spectra ranged between $-2.3 \le \alpha_{turb} \le -1.8$ and did not show any significant trends over $v_{j}$ or $\eta$.
				
				The bulk velocity light curves produced much greater variations over much longer timescales.  The amplitude of the variability was larger for bigger values of $v_{j}$, which is to be expected since larger Lorentz factors will more sharply beam the flux in the direction of flow, and even small changes in that direction will result in major changes in received flux. 
								The slopes of the power spectra produced by these bulk velocity light curves, also given in Table 2, did not show any statistically significant trends over $v_{j}$ or $\eta$.  If such trends exist, it is difficult to find them given the uncertainties in the calculation of PSD slopes ($\approx \pm 0.15$) and our relatively small sample size.  The only apparently significant change in the PSDs is that the $v_j = 0.99c$ cases have the steepest PSD slopes ($-2.8$ and $-2.9$) and the $v_j = 0.995$ case has $\alpha_{bulk} = -2.6$ while all slopes for the slower and faster velocities were in the rather narrow range $-2.4 \le \alpha_{bulk} \le -2.1$.
				
				The slope of the bulk velocity PSD was almost always steeper than that of the turbulence PSD, with an average difference of $\alpha_{bulk} - \alpha_{turb} \approx -0.5$ though those differences had a large range, vanishing for our highest bulk velocity jets ($v_j = 0.998c$ and $v_j =  0.999c$) and with the greatest differences found for the $v_j = 0.99c$ simulations.  With the parameters we used there is a small portion of overlap in the frequencies of the two spectra we computed for each case, and so a cutoff frequency was chosen when the two PSDs were overplotted.  However, in the case of a real blazar, the distinction between these two sources of variability probably will not be as clear, and the different slopes would presumably blend together, smoothing any break in the PSD engendered by different physical mechanisms for variability.   Also, given observational limitations, most actual observations of blazars would typically fall in one or the other of these regimes and so any non-smoothed out break in frequency that could be attributed to the physical mechanisms we have discussed would be unlikely to be detected. In Figure 4, a slight ``hump" can be seen in the high frequency end of the bulk velocity variation PSD just before the match to the turbulent velocity PSDs, particularly for the higher bulk velocity cases.   Similar humps are visible in several of the turbulent jet portions of the simulations as well, and  in those cases are present toward the higher frequency end of those parts of the PSDs before they break to essentially white noise.   However,  these humps    are neither large enough nor consistent enough to indicate any quasi-periodicities in the simulated data.

\subsection{Morphology and Instability}
	
While several of the runs remain completely stable across the entire grid, the majority illustrate the growths of instabilities.  Morphologically, these propagating simulations show similarities to radio jets and lobes seen both in powerful Fanaroff \& Riley (1974) Type II and more modestly powered FR I radio galaxies (Ledlow \& Owen 1996).  The FR I sources are defined as being dominated by emission from the inner half of the source's extent while the FR II sources are dominated by emission from the outer lobes.  These, respectively, typically correspond to sources where the jets have gone unstable and thus the lobes they inflate propagate more slowly, and those where the jets remain stable continue to generate hotspots in the shock complex where they go subsonic, and fill lobes that radiate most intensely near the jet extremities.

Now, with the nominal jet thickness (our effective diameter) we have been using for the variability simulations, the extent of our grid corresponds to only 2400 lt-yr, or sub-galactic, sizes. However, if we rescaled the diameter to a kiloparsec, which is typical for large radio sources (e.g.\ Jeyakumar \& Saikia 2002), the length of the simulation corresponds to 60 kpc, or, considering the opposite lobe as well, some 120 kpc, which is the size of many radio galaxies (the median size of FR IIs is $\sim 300$ kpc, at least for nearby radio galaxies, though it declines with rising redshift; Blundell et al.\ 1999).  Under these circumstances, it seems fair to examine the different morphologies and classify the runs as FR I or FR II, and we have done so in Table 3.  We see that the jets with higher values of $\eta$ and/or $v_j$, and thereby greater thrusts, retain the FR II morphology to the ends of the runs.  Lower densities and lower velocities correspond to weaker jets that are more easily disrupted by perturbations and turn into FR I types.  Several of the runs are undergoing transitions from FR II to FR I types at the ends of the simulations (listed as II/I in the table); sources of this type presumably produce the rare class of hybrid-morphology radio sources, or HYMORS, where one side of the source has a FR I morphology while the other evinces an FR II structure  (e.g., Gopal-Krishna \& Wiita 2000).

	\begin{deluxetable}{c|cccccc}
	\tablenum{3}
	\tablecaption{Morphological Classifications}
	\tablewidth{0pt}
	\tablehead{				
	\colhead{$\eta$} & \colhead{$0.90c$} & \colhead{$0.95c$} & \colhead{$0.99c$} &  \colhead{$0.995c$} & \colhead{$0.998c$} & \colhead{$0.999c$ } 
	}
	\startdata
						 0.1  & II &  II & II & $-$ & $-$ & $-$ \\
						0.01 & I &  I & II & II & II & II \\
						0.00316 & I &  I & I & II/I & II & II \\
						0.001 & \nodata &  \nodata  & I & II/I & II/I &  \nodata \\
	\enddata
	\tablecomments{II/I indicates that the jet shifted from FR II to FR I  before it finished crossing the grid;  $-$ indicates 	the run was of insufficient length to determine the classification.}
	\end{deluxetable}

				\section{Conclusions and Discussion}
	
We have conducted special RHD simulations that combine variations arising from small-scale turbulent flows	with those produced by larger-scale fluctuations in the velocity of propagating jets in order to investigate their suitability for modeling observed blazar variability.  Our simulations of propagating jets using the Athena code have spanned a significant range of velocities for the initial bulk flows ($0.7 \le v_j \le 0.999$) that cover the great majority of the velocities deduced for radio galaxies and blazars (e.g.\ Lister et al.\ 2009).  These flows are light, as is appropriate to radio jets, with $\eta$ values between $10^{-3}$ and $10^{-1}$; although even lower density ratios may be preferred for the best matches to  observations, they are computationally much more difficult to perform and tend to make little difference in the observed morphologies (e.g.\  Krause 2003).  In addition, we have used a recently developed formulation for relativistic turbulence (Zrake \& McFadyen 2013) to model smaller scale turbulent flows along the lines of CW.   

Both the large scale fluctuations in bulk motions and the smaller scale turbulence generate changes in observed emission from the fluid due to changes in the Doppler boosting factors of each fluid element.  We have computed ``total'' emissions for each run by summing those estimated to arise from a slice through our slab jet that is located close to, but upstream from, the primary reconfinement shock.  Those shocks form at several initial jet diameters and stay relatively fixed in location throughout our simulations.  

Our  approach has allowed us to produce light curves and composite power spectra that span roughly five orders of magnitude in time or frequency, something that has not been accomplished previously.  We find that the 	PSDs produced by the turbulent flows are rather consistent, with $\alpha_{turb} \approx -2.0 \pm 0.15$  and nicely overlap those computed from the longer-term variations of most blazars and radio loud AGNs, which normally range between -1.7 and -2.3 (e.g.\ MacLeod et al.\ 2010; Revalski et al.\ 2014; Kasliwal et al.\ 2015).  However, the fluctuations arising from the bulk flow changes tend to produce modestly steeper PSD slopes $\alpha_{bulk} \approx -2.5 \pm 0.2$  
 that match only a small subset of observed AGNs (Mushotzky et al.\ 2011; Carini \& Ryle  2012; Kasliwal et al.\ 2015).   We see that in some cases, particularly for the highest bulk velocities of $0.998c$ and $0.999c$, the PSD slopes and the turbulence slopes are very close and thus yield a very smooth match that could well correspond to the typical observations.  With the parameters of our simulations the bulk fluctuations are typically produced on timescales of years to many decades while the modeled turbulent variations span several days to years.  The best observations to date typically span hours to a few years, so the general agreement with the shallower slopes produced by the turbulent fluctuations is reasonable. We expect that only with observations extending over much longer time spans might we discover many powerful radio-loud AGNs with the steeper PSD slopes produced by the majority of the bulk Doppler induced variations.  It will be necessary to extend and improve these simulations to produce models that can be compared in detail with observations.

				Several important simplifications mean that this work can only be considered a step toward modeling the full variability of radio sources over such wide temporal intervals.  Clearly the 2D nature of our simulations instead of a full 3D means that our results can only be taken as approximations to the actual fluctuations a jet would exhibit during propatation.   A slab jet of given ``power''  will advance faster than than a 3D simulation of a cylindrical jet when the slab thickness is taken equivalent to the jet diameter, $d$ for that computation.  This is because as the slab jet spreads out  the overall thrust will scale closer to $d^{-1}$ rather than the $d^{-2}$ it would for a widening cylindrical jet.  Once the jets go unstable then this effect becomes important and the overall advance of the lobe will indeed be faster than it would be in a 3D simulation.  But because the stable jet itself does not expand dramatically for the length of our simulations, this effect should not be substantial as far as the light curve calculations are concerned.  It is worth noting that analogs exist for modes normally present in 3D jets in these 2D simulations (e.g., Hardee \& Norman 1988, 1990).  The $m=1$ pinch mode is fairly straightforward to map to three dimensions, as it  results in the formation of conical recollimation shocks instead of roughly X-shaped ones.  The $m=2$ sinusoidal perturbations that are present in our simulations and contribute greatly to the bulk velocity variations we have mapped can be seen to correspond to helical perturbations in 3D jets (e.g., Hardee \& Norman 1988).

	A second major simplification in our computations involves their restriction to RHD flows as opposed to RMHD ones.  In the absence of magnetic field strengths and orientations we had to make the crude assumption that the rest-frame emissivity scaled with the relativistic plasma density.  While this has been done in a large number of non-relativistic and relativistic previous works (e.g., Hooda et al.\ 1994; Mart{\'i} et al.\ 1997; Mioduszewski et al.\ 1997), state-of-the art simulations do now incorporate the magnetic fields  both in computing the speeds and structures of  jets and in estimating emissivities (e.g.\  Komissarov 1999; Gaibler et al.\ 2009; Marscher 2014).   Although the Athena code  is optimized for MHD computations, and we experimented with them,  employing the MHD module along with the RHD one requires substantially more computer memory as well as much more clock time to complete each timestep.  Because we currently lack access to functioning clusters, the only way we could compute reasonably well-resolved simulations over the substantial problem times  we needed so as to obtain lengthy light curves, while only employing single workstations, was to stay in two dimensions instead of three dimensions and to not turn on the MHD capabilities.  
				
				The work of Marscher (2014) is notable in that the simulated flux was integrated over up to 1140 turbulent cells in his 3D mesh, as a shock's propagation through a Mach disk in the reconfinement shock region is followed.  This is certainly better in principle than our approach since our light curves only used flux data calculated from one column (typically 20) of cells.  His code also incorporates variations in the magnetic field strengths and orientations and thus allows for much more physically motivated estimation of the flux from each cell as a function of time and this means that polarization properties of the flux can be followed and compared with observations.  Even more importantly, his approach allows  the computation of theoretical fluxes over a wide range of the electromagnetic spectrum (radio to gamma-ray). This produces reasonably detailed comparisons to the light curves of the limited number of blazars where simultaneous data has been obtained through multi-wavelength campaigns.  On the other hand, his approach requires the assumption of a particular set of initial bulk velocities for his cells whereas our approach naturally produces them from our propagating jet simulation.

Clearly we can improve on the current work by relaxing each of the major simplifications we have used so far when we have access to substantially greater computational resources.  The first step would be to convert our 2D simulations to 3D ones and then increase the number of zones whose fluxes are summed over to produce light curves.  This improvement would also involve using differences in velocities between the bulk simulation zones to provide more realistic maximum sub-grid turbulent velocities that could vary from zone to zone. Ideally, the fluxes from all zones in the entire grid would be summed over, although this would be very challenging computationally.  The second step would be to combine the MHD and RHD modules to be able to compute the dynamical effects of magnetic fields on the 3D jet's propagation as well as to produce substantially better estimates of the emissivity.  We plan on performing  this additional work, as adequate computational power should be available to us in the near future.

\end{document}